\documentclass[prd,superscriptaddress,showpacs,nofootinbib,amsmath,amssymb, twocolumn,aps,10pt]{revtex4-1}

\usepackage{bm}
\usepackage{amsfonts}
\usepackage{latexsym}
\usepackage[latin1]{inputenc}
\usepackage{graphicx}
\usepackage{amsmath}
\usepackage{palatino}
\usepackage{mathpazo}
\usepackage{booktabs}
\usepackage{dcolumn}

\def \nn  {\nonumber}

\def\jnl@style{\it}
\def\aaref@jnl#1{{\jnl@style#1}}
\def\aaref@jnl#1{{\jnl@style#1}}

\def\aj{\aaref@jnl{AJ}}                   % Astronomical Journal
\def\apj{\aaref@jnl{ApJ}}                 % Astrophysical Journal
\def\apjl{\aaref@jnl{ApJ}}                % Astrophysical Journal, Letters
\def\apjs{\aaref@jnl{ApJS}}               % Astrophysical Journal, Supplement
\def\apss{\aaref@jnl{Ap\&SS}}             % Astrophysics and Space Science
\def\aap{\aaref@jnl{A\&A}}                % Astronomy and Astrophysics
\def\aapr{\aaref@jnl{A\&A~Rev.}}          % Astronomy and Astrophysics Reviews
\def\aaps{\aaref@jnl{A\&AS}}              % Astronomy and Astrophysics, Supplement
\def\mnras{\aaref@jnl{MNRAS}}             % Monthly Notices of the RAS
\def\prd{\aaref@jnl{Phys.~Rev.~D}}        % Physical Review D
\def\prl{\aaref@jnl{Phys.~Rev.~Lett.}}    % Physical Review Letters
\def\qjras{\aaref@jnl{QJRAS}}             % Quarterly Journal of the RAS
\def\skytel{\aaref@jnl{S\&T}}             % Sky and Telescope
\def\ssr{\aaref@jnl{Space~Sci.~Rev.}}     % Space Science Reviews
\def\zap{\aaref@jnl{ZAp}}                 % Zeitschrift fuer Astrophysik
\def\nat{\aaref@jnl{Nature}}              % Nature
\def\aplett{\aaref@jnl{Astrophys.~Lett.}} % Astrophysics Letters
\def\apspr{\aaref@jnl{Astrophys.~Space~Phys.~Res.}} % Astrophysics Space Physics Research
\def\physrep{\aaref@jnl{Phys.~Rep.}}      % Physics Reports
\def\physscr{\aaref@jnl{Phys.~Scr}}       % Physica Scripta
\def\commat{\aaref@jnl{Comm.~Math.~Phys.}}		% Communications in Mathematical Physics
\def\science{\aaref@jnl{Science}}		% Science
\def\cqg{\aaref@jnl{Classical Quant.~Grav.}}		% Classical and Quantum Gravity
\def\jpcs{\aaref@jnl{JPCS}}					% Journal of Physics Conference Series
\def\ijmpd{\aaref@jnl{Int.~J.~Mod.~Phys.~D}}			% International Journal of Modern Physics D
\def\grg{\aaref@jnl{Gen.~Relat.~Gravit.}}		% General Relativity and Gravitation
\def\rpp{\aaref@jnl{Rep.~Prog.~Phys.}}		% Reports on Progress in Physics

\allowdisplaybreaks[1]
\DeclareMathOperator{\sgn}{sgn}

\begin{document}

\title{Gravitational wave asteroseismology with fast rotating neutron stars}

\author{Erich Gaertig}
%\email{gaertig@tat.physik.uni-tuebingen.de}
\affiliation{Theoretical Astrophysics, Eberhard-Karls University of T\"ubingen, T\"ubingen 72076, Germany}
\author{Kostas D. Kokkotas} % \email{kokkotas@auth.gr}
\affiliation{Theoretical Astrophysics, Eberhard-Karls University of T\"ubingen, T\"ubingen 72076, Germany}
\affiliation{Department of Physics, Aristotle University of Thessaloniki, Thessaloniki 54124, Greece}
%%%%%%%%%%%%%%%%%%%%%%%%%%%%%%%%%%%%  DATE  %%%%%%%%%%%%%%%%%%%%%%%%%%%%%%%%%%%%
\date{\today}

\begin{abstract}
We investigate damping and growth times of the quadrupolar $f$-mode for rapidly rotating stars and a variety of different polytropic equations of state in the Cowling approximation. 
This is the first study of the damping/growth time of this type of oscillations for fast rotating neutron stars in a relativistic treatment where the spacetime degrees of freedom of the perturbations are neglected.
We use these frequencies and damping/growth times to create robust empirical formulae which can be used for gravitational wave asteroseismology.
The estimation of the damping/growth time is based on the quadrupole formula and our results agree very well with Newtonian ones in the appropriate limit. 
\end{abstract}

%%%%%%%%%%%%%%%%%%%%%%%%%%%%%  PACS  %%%%%%%%%%%%%%%%%%%%%%%%%%%%%%%%%%%%%%%%%%%

%%%%%%%%%%%%%%%%%%%%%%%%%%%%%  MAKETITLE  %%%%%%%%%%%%%%%%%%%%%%%%%%%%%%%%%%%%%%
\maketitle

%=====================================================================
\section{Introduction}
\label{sec:introduction}
%=====================================================================

During the birth of a proto-neutron star or the merging of two older compact stars, violent non-radial oscillations
may be excited, resulting in the emission of significant amounts of gravitational radiation \cite{Andersson:2011lr}.
The detection of gravitational waves from oscillating neutron stars will allow the study of their interior, in the same way as helioseismology provides information about the interior of the Sun. It is expected that the identification of specific pulsation frequencies in the observational data will reveal the true properties of matter at densities that cannot be probed today by any other experiment. In this paper, we present new empirical relationships for mode-frequencies and damping times of the quadrupolar $f$-mode for rapidly rotating neutron stars, extending previous studies which deal with the non-rotating case \cite{1996PhRvL..77.4134A,1998MNRAS.299.1059A,2001MNRAS.320..307K}.

%Thus, there is significant interest in this problem also from the high-energy-physics community since the discovery of the true properties of matter at extremely high energies will allow significant advances in theoretical physics and possibly new discoveries and applications will follow.

These original suggestions about gravitational wave asteroseismology have been supported by many complementary works which studied specific features of oscillation spectra for various compact objects, such as typical neutron stars \cite{1999MNRAS.310..797B,2002PhRvD..65b4010S,2004PhRvD..70l4015B,2005PhRvL..95o1101T,2006PhRvD..74l4025T,2007PhRvD..76d3003C,2007MNRAS.381..151W,Lau:2010lr}, but also for strange \cite{2003PhRvD..68b4019S,2004PhRvD..69h4008S,Benhar:2007lr} or superfluid stars \cite{2001PhRvL..87x1101A,2002PhRvD..66j4002A}. More recently, it has also been suggested that one may use asteroseismology to find the imprints of scalar or even vector components of gravity \cite{2004PhRvD..70h4026S,2005PhRvD..71l4038S,2009PhRvD..79f4033S,2009PhRvD..80f4035S}.
It should be noted here that all previous studies  \cite{Andersson:2011lr,1996PhRvL..77.4134A,1998MNRAS.299.1059A,2001MNRAS.320..307K,1999MNRAS.310..797B,2002PhRvD..65b4010S,2004PhRvD..70l4015B,2005PhRvL..95o1101T,2006PhRvD..74l4025T,2007PhRvD..76d3003C,2007MNRAS.381..151W,Lau:2010lr,2003PhRvD..68b4019S,2004PhRvD..69h4008S,Benhar:2007lr,2001PhRvL..87x1101A,2002PhRvD..66j4002A,2004PhRvD..70h4026S,2005PhRvD..71l4038S,2009PhRvD..79f4033S,2009PhRvD..80f4035S} have been performed for non-rotating relativistic stars. The treatment of rotation was always a problem in general relativity and thus the majority of the studies for the oscillation spectra of fast-rotating compact stars was done mainly in Newtonian theory which gives only qualitative answers. 

%Stellar oscillations may become unstable in the presence of rotation. 
 Since stellar oscillations may become unstable in the presence of rotation, there was  an increased interest during the last decade or so to study the dynamics of rotating stars, also thanks to the discovery of the $r$- and $w$-mode instability \cite{1998ApJ...502..708A,1998ApJ...502..714F,1998PhRvL..80.4843L,1999ApJ...510..846A,Kokkotas:2004lr}. 
Still, the majority of these studies have been performed in Newtonian theory \cite{1999PhRvD..59d4009L,1999ApJ...521..764L,1999PhRvD..60f4006L} while there are only a few works in which GR has been used; mainly in the so called {\em slow-rotation approximation}. The slow-rotation approximation was successfully applied to study various aspects of the $r$-mode instability \cite{2001IJMPD..10..381A,2001MNRAS.328..678R}, effects of uniform and differential rotation on the oscillation spectrum \cite{Ferrari:2004qy,2007PhRvD..75f4019S,2008PhRvD..77b4029P} and on the crustal modes \cite{2008MNRAS.384.1711V}.

As it has been originally suggested  by Chandrasekhar \cite{1970PhRvL..24..611C} and verified by Friedman \& Schutz \cite{1978ApJ...221..937F,1978ApJ...222..281F}  certain non-axisymmetric pulsation modes may grow exponentially in rotating stars; this is due to the emission of gravitational waves and is called  CFS instability. Exploring this type of instability in rapidly rotating stars turned out to be very difficult. In linear perturbation theory for example, rapid rotation was never treated properly until recently; almost all formulations of the relevant perturbation equations were prone to numerical instabilities either at the surface or along the rotation axis of the neutron star. Thus, it was not surpising that the first results for the oscillations of rapidly rotating stars were derived  using evolutions of the non-linear equations \cite{2004MNRAS.352.1089S,2006MNRAS.368.1609D,2006PhRvD..74f4024K,2008PhRvD..77f4019K}. Still all these studies were purely axisymmetric and thus the effects of rotation on the spectra was present only for very high rotation rates. Rotational instabilities are driven by non-axisymmetric modes and thus these first 2D calculations where not of much use for their study.

In the last two years there was significant progress in the study of non-axisymmetric perturbations of rapidly rotating neutron stars. For the first time it was possible to calculate in GR the oscillation spectra of fast rotating relativistic stars by using the linearized form of the fluid equations.  Thus the effect of fast rotation on  $f$- and $r$-modes has been demonstrated  while the  critical points for the onset of the $f$-mode (CFS) instability have been derived \cite{2008PhRvD..78f4063G}. In addition it has been demonstrated that there is a way to derive empirical relations connecting the oscillation frequencies with the rotation of the stars. This study has been recently extended  to $g$-modes \cite{2009PhRvD..80f4026G} and even more recently has been expanded to study the oscillation spectra of fast and differentially rotating neutron stars \cite{2010PhRvD..81h4019K}.

It should be noted that the previous results have been derived using the so-called Cowling approximation where the spacetime is assumed to be frozen. This approximation is very good for the estimation of the spectra of $r$- and $g$-modes but it gives only qualitatively good results for the $f$-mode. Moreover, using non-linear codes it became possible for the first time to study the complete problem \cite{PhysRevD.81.084055}, i.~e.~the non-axisymmetric stellar oscillations of fast rotating stars without the constraints of the Cowling approximation. The results are in qualitative agreement with those found in \cite{2008PhRvD..78f4063G} and for the critical point for the onset of the $f$-mode instability with the studies presented in \cite{1998ApJ...492..301S}.

The next step for gravitational wave asteroseismology is to use additional information about the damping times to construct model-independent relations which allow for a robust determination of stellar key parameters. The damping time $\tau$ of the potentially CFS-unstable branch in the high rotation regime for example can be approximated very accurately by using a simple relation of the form $\tau_0/\tau = \sgn{(\sigma_i)}\,0.256(\sigma_i/\sigma_0)^4$, where $\tau_0$ and $\sigma_0$ are the damping time and mode frequency of the nonrotating model respectively, $\sgn{(x)}$ is the signum function and $\sigma_i$ is the actual mode frequency in the inertial frame.

The structure of the paper is as follows. In Section \ref{sec:problem_setup} we give an essential overview about our method of computing mode-frequencies and damping times of the $f$-mode. We then show the results of our simulations in Section \ref{sec:empirical_relations}, where we present empirical relations which can potentially be used to estimate masses, radii and angular frequencies of rapidly rotating neutron stars. A more elaborate discussion about the numerical procedure, the equations of state and background models used in this study as well as a consistency check can be found in the Appendix.
%=====================================================================
\section{Problem Setup}
\label{sec:problem_setup}
%=====================================================================

Mode frequencies and damping times of neutron star oscillations can be calculated in two different ways. In a time-independent boundary-value formulation of the perturbation equations, they are directly obtained from the real and the imaginary part of the complex eigenfrequencies. In a time-dependent evolution problem on the other hand, both these quantities have to be computed in post-processing routines. The frequency of specific fluid modes is obtained by Fourier-transforming the time-series at different points inside the star into the frequency domain and correctly identifying the corresponding peaks in the power spectrum, see \cite{2008PhRvD..78f4063G} for a detailed description of this method.

Concerning the damping time, one has to calculate both the energy that is contained within a mode as well as the energy loss due to gravitational wave emission which in our case is done via the quadrupole formula, see e.~g.~\cite{1999ApJ...521..764L} for an application of this formalism to $r$-modes. A different procedure is to use the behaviour of metric perturbations at future null infinity to derive a gauge-invariant expression for the luminosity, see \cite{Lockitch:2003lr}. However, since we will work in the Cowling-approximation, the quadrupole formalism is utilized in this study.

The energy of a mode in a comoving frame is then given by
\begin{equation}
\label{eq:modeEnergy}
E = \frac{1}{2}\int\left[\rho\delta u^{a} \delta u^{\ast}_{a} + \left(\frac{\delta p}{\rho} + \delta \Phi\right)\delta\rho^{\ast} \right]d^3x\,,
\end{equation}
where $\rho$ is the rest-mass density and $\delta\rho$ its corresponding perturbation, $\delta p$, $\delta\Phi$ the perturbations of pressure and gravitational potential respectively. On the other hand, the quadrupole formula for the emission of gravitational radiation gives
\begin{equation}
\label{eq:energyLoss}
\frac{dE}{dt} = - \sigma_i(\sigma_i + m\Omega)\sum_{l \geq 2}N_l \sigma_i^{2l}(|\delta D_{lm}|^2 + |\delta J_{lm}|^2)\,,
\end{equation}
where
\begin{equation}
\label{eq:couplingConstant}
N_l = \frac{4\pi\, G}{c^{2l + 1}}\frac{(l+1)(l+2)}{l(l-1)[(2l+1)!!]^2}
\end{equation}
is the coupling constant for spherical mode number $l$, $\sigma_i$ the mode frequency in the inertial frame, $\Omega$ the angular velocity of the neutron star and where $\delta D_{lm}, \delta J_{lm}$ are the mass- and the current-multipole moments respectively. The damping time is then computed by
\begin{equation}
\label{eq:dampingTime}
\frac{1}{\tau_{gr}} = -\frac{1}{2E}\frac{dE}{dt}\,.
\end{equation}
Several remarks are now worth considering. First, as already mentioned we are working in the Cowling-approximation which means that the spacetime is kept fixed during the time-evolution, i.~e.~$\delta\Phi = 0$ in equation \eqref{eq:modeEnergy}. Second, we will focus on the nonaxisymmetric $l = |m| = 2$ fundamental mode since in general it has the smallest growth-time and is therefore more relevant in real astrophysical scenarios. Higher order modes typically not only grow on larger timescales but are also damped stronger by various dissipative effects. In the case of pressure modes, the emission of gravitational waves is to a great extent due to the mass-quadrupole moment and we will neglect the current quadrupole moment $\delta J_{22}$. Taking these comments into account, equations \eqref{eq:modeEnergy} and \eqref{eq:energyLoss} adjusted for $f$-modes in the Cowling-approximation, read
\begin{equation}
\label{eq:equationsForFModesInCowling_I}
E = \frac{1}{2}\int\left[\rho\delta u^{a} \delta u^{\ast}_{a} + \frac{\delta p}{\rho}\delta\rho^{\ast} \right]d^3x
\end{equation}
and
\begin{equation}
\label{eq:equationsForFModesInCowling_II}
\frac{dE}{dt} = - \sigma_i(\sigma_i + m\Omega)N_2 \sigma_i^{4}|\delta D_{22}|^2
\end{equation}
with
\begin{equation}
\label{eq:equationsForFModesInCowling_III}
D_{22} = \int\delta\rho\,r^2 Y^{\ast}_{22}d^3x
\end{equation}
as mass-quadrupole moment.

\begin{figure}[ht!] 
\centering
\includegraphics[width=0.48\textwidth]{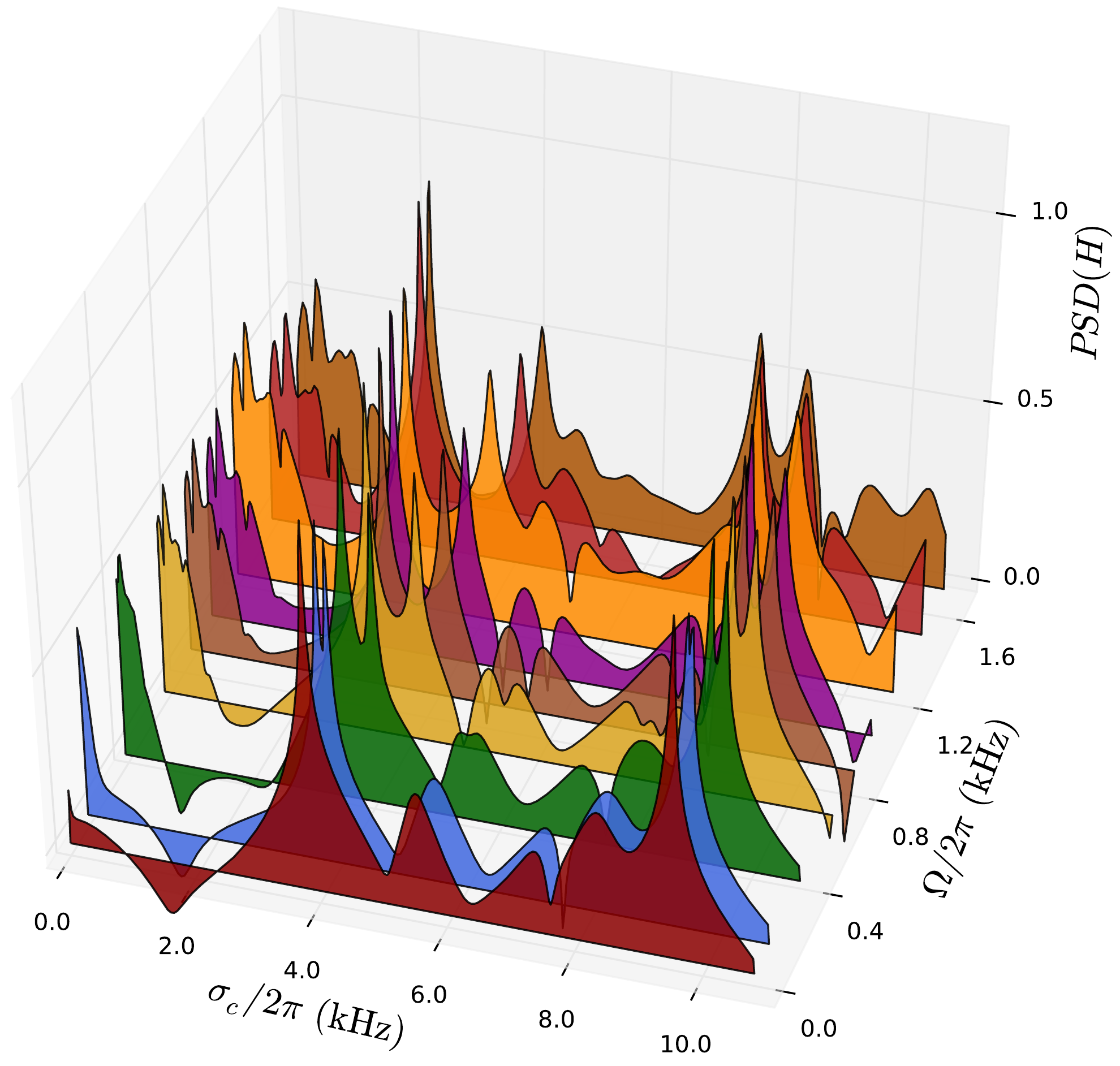}
\caption{Splitting of the power spectral density (normalized units) for non-axisymmetric $|m| = 2$-modes in a comoving reference frame with corresponding mode-frequency $\sigma_c$. The mass-shedding limit for this particular sequence is reached for $\Omega/2\pi = 2.18$ kHz.}
\label{fig:modeSplitting_eosC}
\end{figure}

The damping time $\tau_{gr}$ depends crucially on how the mode frequencies of non-axisymmetric perturbations change with rotation rate. While degenerated in the non-rotating limit, the frequencies of modes with the same spherical mode number $l$ but opposite azimuthal index $m=\pm |m|$, i.~e.~co- and counterrotating modes, diverge. Figure \ref{fig:modeSplitting_eosC} shows an example of this behaviour for a certain sequence of equilibrium models with increasing angular frequency $\Omega/2\pi$ in a system comoving with the star. The power spectral density of the pressure perturbation variable $H$ is taken at an arbitrary point inside the neutron star (typically $s = t = 0.5$; for a description of the computational domain and the coordinates used there, see Appendix \ref{sec:numerical_procedure}) and colour-coded for the different models. In the nonrotating limit, one can identify various peaks with the strongest and sharpest ones located at $\sigma_1/2\pi = 3.837$ kHz and at $\sigma_2/2\pi = 9.432$ kHz.

%%%%%%%%%%%%% Figure %%%%%%%%%%%%%%%%%%%%%%%%
\noindent
%%%%%%%%%%%%% Figure %%%%%%%%%%%%%%%%%%%%%%%%
Inspection of the corresponding eigenfunctions shows that the peak at $\nu_1$ belongs to the quadrupolar $f$-mode while $\nu_2$ matches its first overtone, the $^2p_1$-mode with an additional node in radial direction. Increasing the angular velocity leads to a splitting of nonaxisymmetric perturbations which can most clearly be seen for the two modes mentioned. Other peaks in Figure \ref{fig:modeSplitting_eosC} split as well, however due to broader edges this behaviour is harder to follow.

The imaginary part of the complex mode frequency, which is approximated by equation \eqref{eq:dampingTime}, controls the exponential damping or growing of non-axisymmetric perturbations. For nonrotating stars it is $E > 0$ and $dE/dt < 0$, see equations \eqref{eq:modeEnergy} and \eqref{eq:energyLoss}. The imaginary part of the mode frequency then becomes positive, indicating a damped oscillation. The perturbations remain damped as long as $\sigma_i(\sigma_i + m\Omega) > 0$, that is counter-rotating modes in the comoving frame are still counter-rotating in the inertial frame. This behaviour changes once the pattern speed of the mode is matched by the angular velocity. In this case, $dE/dt = 0$ and there is no loss of energy due to gravitational radiation. Finally, for background configurations that allow mode frequencies beyond the zero-frequency limit in the inertial frame it is $dE/dt > 0$ and the oscillation is exponentially growing on a timescale given by $\tau_{gr}$.

These are the astrophysically most interesting cases since the oscillation is unstable in this regime, emitting significant amounts of gravitational radiation. While damping times have already been computed for nonrotating stars \cite{1983ApJS...53...73L,1998MNRAS.299.1059A,1999MNRAS.310..797B,2001MNRAS.320..307K,2004PhRvD..70l4015B} as well as in the relativistic slow-rotation approximation \cite{Ferrari:2007lr}, so far there are no numerical simulations for calculating damping times of rapidly rotating relativistic models. However, in Newtonian theory the quadrupole formula has been used successfully for computing damping times of rotating polytropes \cite{Ipser:1990lr,Ipser:1991fk} and recently also for gravitational-wave extraction of rotating superfluid stars \cite{Passamonti:2010lr}. 

%=====================================================================
\section{Asteroseismology}
\label{sec:empirical_relations}
%=====================================================================

%=====================================================================
\subsection{Frequencies of Co- and Counterrotating Modes}
\label{ssec:frequency_co_and_counter}
%=====================================================================

%%%%%%%%%%%%% Figure %%%%%%%%%%%%%%%%%%%%%%%%
\begin{figure*}[htp!]
\centering
\includegraphics[width=0.95\textwidth]{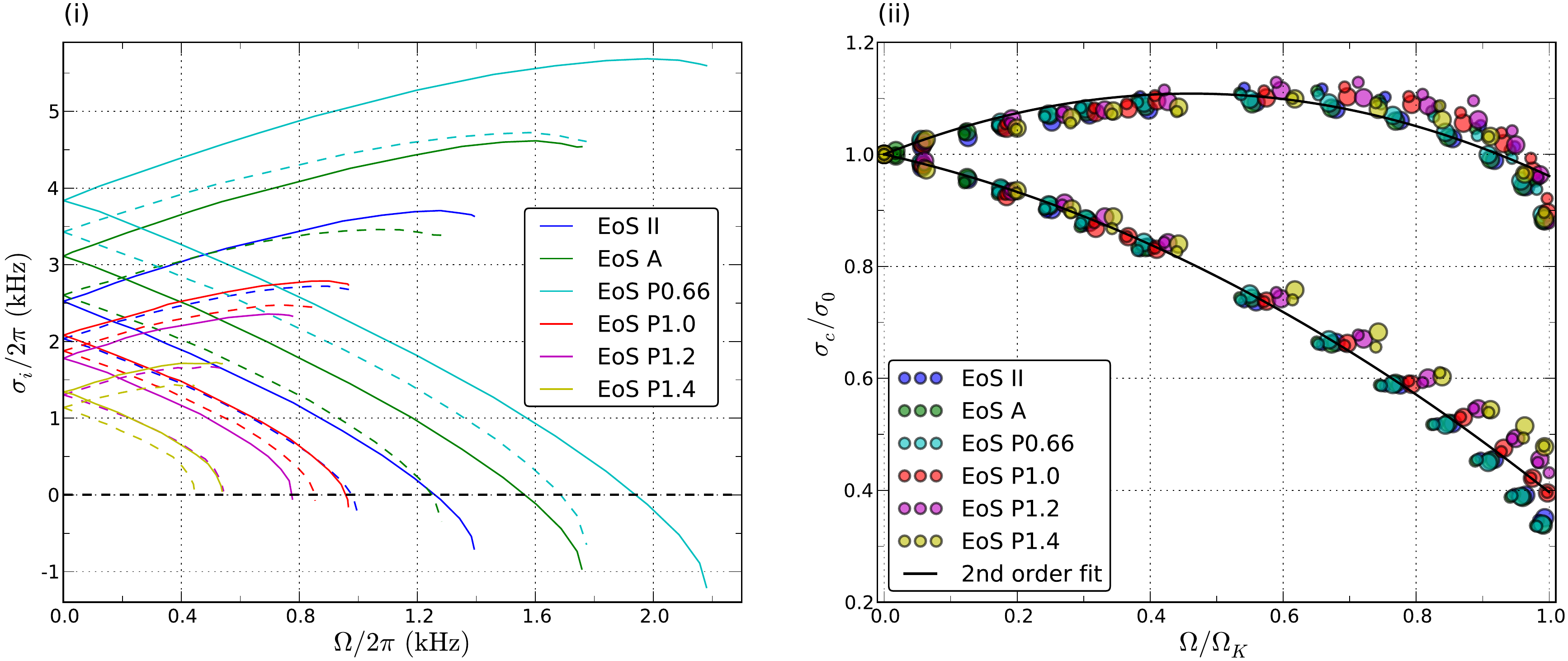}
\caption{Co- and counterrotating branches of the different polytropic EoS studied in this paper. {\it Panel (i):} Mode frequencies in the inertial frame; the solid line of each EoS depicts the more compact sequence while the dashed line traces the less compact configurations. {\it Panel (ii):} Normalized mode frequencies and fitting curve in the comoving frame; the larger circles represent again the more compact models while the small circles stand for the less compact ones. $\sigma_0$ is the frequency in the nonrotating limit, $\Omega_K$ represents the Kepler-limit.}
\label{fig:splittingCombined}
\end{figure*}
\noindent
%%%%%%%%%%%%% Figure %%%%%%%%%%%%%%%%%%%%%%%%

In \cite{2008PhRvD..78f4063G} we already presented results about the splitting of the fundamental mode in a coordinate frame comoving with the star. The conclusions there suggested, that while in the inertial frame the exact details of the $f$-mode splitting depend on the particular background model and the given equation of state, when properly normalized in the comoving frame, one can actually construct a model-independent relation between mode frequency and rotation rate.   

In parts, this is due to the following reason. In the inertial frame there is a clear cut between configurations that allow for potentially CFS-unstable models and configurations which never become CFS-unstable. If one were to fit for example all counter-rotating modes with just one fitting curve, either all models will become unstable at some point or no model at all. Clearly, this would be implausible for a proper fitting.

In the comoving frame on the other hand, the situation is different. There, the two branches of counter- and corotating modes never reach the zero-frequency limit even remotely and this is due to the fact that the absolute value of the splitting between the $m = |m|$ and the $m = -|m|$ branches is smaller there. The relation between mode frequencies $\sigma_i$ in the inertial frame and $\sigma_c$ in the comoving frame is given by

\begin{equation}
\label{eq:inertialAndComovingFrequencyTransformation}
\sigma_i = \sigma_c - m\Omega\,.
\end{equation}
It then follows from equation \eqref{eq:inertialAndComovingFrequencyTransformation}, that the frequency separation $\Delta$ between the two branches transforms according to
\begin{equation}
\label{eq:gapTransformation}
\Delta_i = \Delta_c + 2|m|\Omega\,,
\end{equation}
that is the separation is always smaller in the comoving system and it is actually the corotating branch that has the lower frequencies in this frame, see e.~g.~\cite{2008PhRvD..78f4063G,2002MNRAS.334..933J}.

Figure \ref{fig:splittingCombined} shows the results of our simulations regarding the mode frequencies and as expected, they show a large variety in the inertial frame. Depending on the actual configuration and equation of state, some models become unstable before reaching the mass-shedding limit, some are only marginally unstable and some remain stable even at the Kepler frequency.

Despite this apparent diversity in the inertial frame, the two branches can very well be fitted with a second order polynomial in the comoving frame, see Panel $(ii)$ of Figure \ref{fig:splittingCombined}. There, the mode frequency is normalized by its value in the nonrotating limit $\sigma_0$ while the angular velocity is prescribed in units of the Kepler frequency $\Omega_K$. As one can see, the fitting describes the overall behaviour of the mode frequencies very well; only for values close to the mass-shedding limit the various curves for the different EoS show larger deviations from the quadratic fit. One should also keep in mind that the order of co- and counterrotating modes is reversed in the comoving frame, that is while unstable modes have lower frequencies in the inertial frame, they represent the high-frequency branch in the comoving frame and vice versa.

Based on the data of our simulations, we propose the following relationships for the quadratic fitting polynomials. It is
\begin{equation}
\label{eq:sigmaStableBranch}
\frac{\sigma^{s}_c}{\sigma_0} = 1.0 - 0.27\left(\frac{\Omega}{\Omega_K}\right) - 0.34 \left(\frac{\Omega}{\Omega_K}\right)^2
\end{equation}
for the always stable ($m = -2$) and
\begin{equation}
\label{eq:sigmaUnstableBranch}
\frac{\sigma^{u}_c}{\sigma_0} = 1.0 + 0.47\left(\frac{\Omega}{\Omega_K}\right) - 0.51 \left(\frac{\Omega}{\Omega_K}\right)^2
\end{equation}
for the potentially unstable ($m = 2$) branch. This also agrees very well with our previous findings for a more limited set of equilibrium configurations and equations of state in \cite{2008PhRvD..78f4063G}.

In addition, an auxiliary condition is needed that connects the mode frequency in the nonrotating limit $\sigma_0$ with fundamental stellar parameters. It is well known, that for the $f$-mode $\sigma_0$ scales with the mean density $(M/R^3)^{1/2}$, see e.~g.~\cite{1998MNRAS.299.1059A,2001MNRAS.320..307K}, where fitting coefficients based on a variety of realistic EoS are provided.

\begin{figure}[htp!]
\centering
\includegraphics[width=0.48\textwidth]{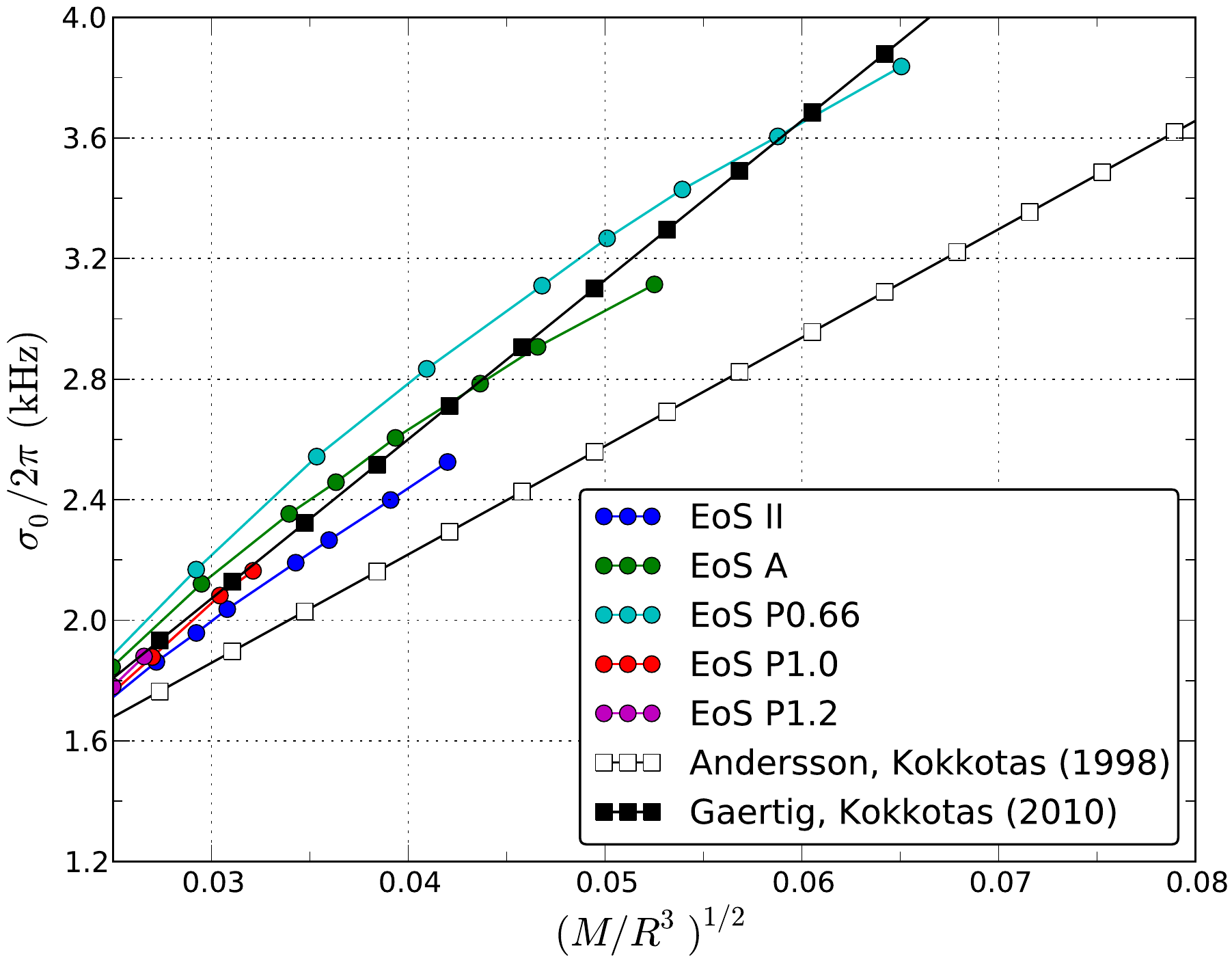}
\caption{Scaling of the $f$-mode frequency with mean density in the nonrotating limit. $M$ and $R$ are given in km, $\sigma_0$ in kHz.}
\label{fig:nonrotFrequency}
\end{figure}

We repeated this calculation in the time-domain and with our set of equation of states and the results are depicted in Figure \ref{fig:nonrotFrequency}. In order to better compare them with the findings in \cite{1998MNRAS.299.1059A}, we picked out the very same range of mean densities. Depending on the particular EoS, it may happen that certain configurations never reach the range of mean densities depicted in Figure \ref{fig:nonrotFrequency}; for example the very stiff EoS P1.4 is completely absent.

%%%%%%%%%%%%% Figure %%%%%%%%%%%%%%%%%%%%%%%%
\noindent
%%%%%%%%%%%%% Figure %%%%%%%%%%%%%%%%%%%%%%%%
The frequencies computed with our approach generally tend towards larger values. This can very well be understood with the Cowling-approximation which is used to simplify the time-evolution equations, see Appendix \ref{sec:numerical_procedure}. The freezing of all metric perturbations systematically overestimates pressure-mode frequencies though this effect becomes less pronounced for higher order modes, see also \cite{PhysRevD.81.084055} and references therein. Another difference is the use of polytropic equations of state in contrast to tabulated EoS utilized in \cite{1998MNRAS.299.1059A}; this also has an effect on the mode frequencies and preliminary studies show that this might be an even stronger restriction than the Cowling-approximation \cite{Kruger:fk}.

Nevertheless, it is still possible to fit the frequencies very well with a linear dependence in the mean density and we find
\begin{equation}
\label{eq:sigmaZeroNonrot}
\frac{1}{2\pi}\sigma_0\,\mathrm{(kHz)} = 0.498 + 2.418 \left(\frac{\bar M}{\bar R^3}\right)^{1/2}\,,
\end{equation}
where we introduced the dimensionless variables
\begin{equation}
\label{eq:M_and_R_Normalized}
\bar M = \frac{M}{\mathrm{1.4}\, M_\odot}\quad\mathrm{and}\quad\bar R = \frac{R}{\mathrm{10\,km}}\,.
\end{equation}
%One should keep in mind that the difference to the values presented in \cite{1998MNRAS.299.1059A} which were computed in full general relativity, is to a great extent due to the Cowling approximation used here.

%=====================================================================
\subsection{Damping Times of Co- and Counterrotating Modes}
\label{ssec:dampTime_co_and_counter}
%=====================================================================

A similar procedure can be applied to the damping times of the two fundamental mode branches. However, in order to find a model-independent relation, we cannot use the angular velocity $\Omega$ directly as a measure of the rotation rate as it was done in Panel $(ii)$ of Figure \ref{fig:splittingCombined}. The reason for this is, that the damping time changes its sign when a particular fundamental mode eventually becomes unstable, see the discussion in Section \ref{sec:problem_setup}. A negative damping time signals an exponential growth instead of a damped oscillation. If one were to fit the various damping times of the counterrotating modes as function of angular velocity with just one fitting curve, then again either all models would become unstable at some point or no model at all; see also the discussion in Section \ref{ssec:frequency_co_and_counter} where we discussed a similar effect for mode frequencies in the inertial frame.

\begin{figure*}[htp!]
\centering
\includegraphics[width=0.67\textwidth]{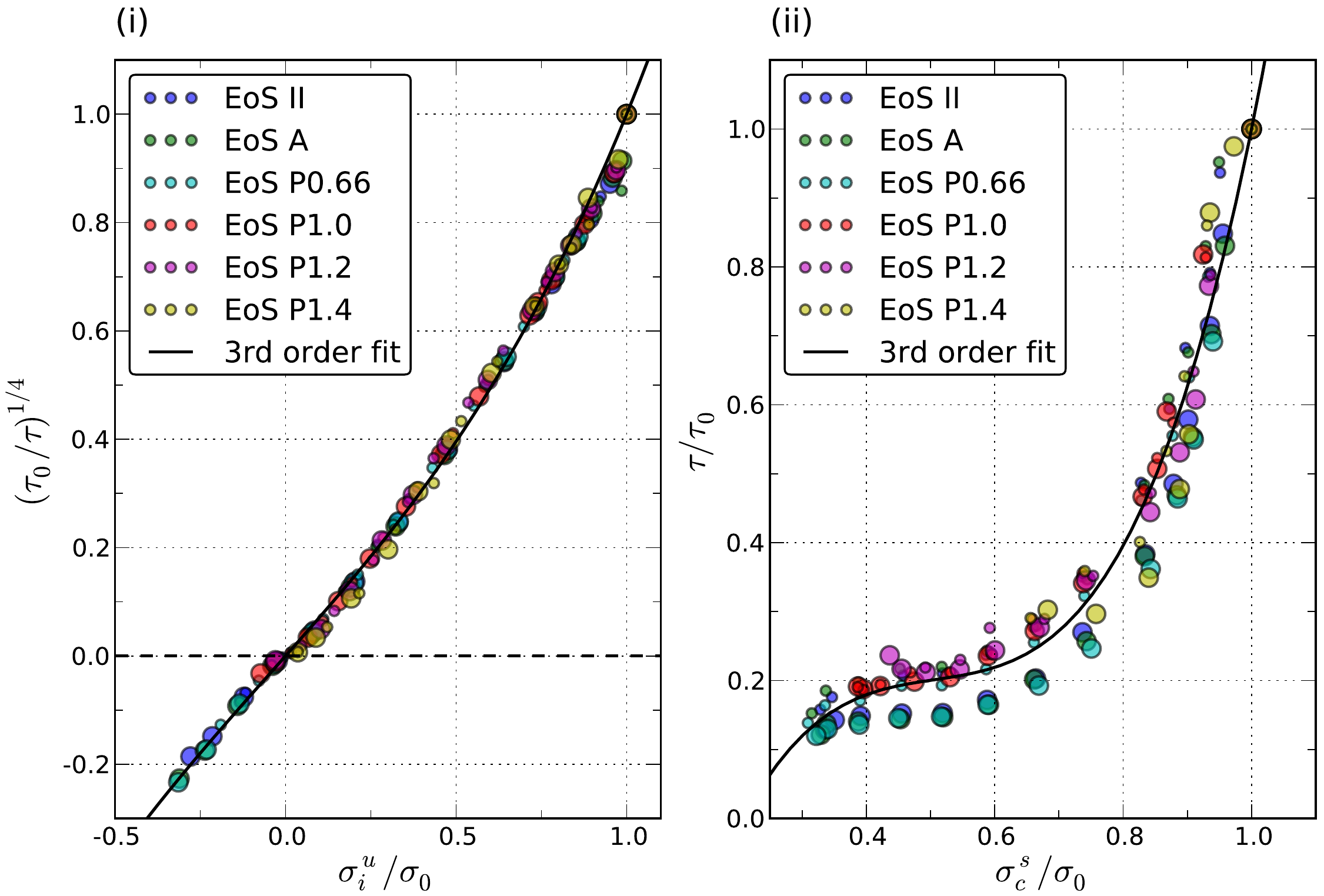}
\caption{Model-independent relations for the damping time. Larger circles represent the more compact models while small circles stand for the less compact configurations of each EoS. {\it Panel (i):} Damping times for the counterrotating branch, $\sigma^{u}_i$ is the mode frequency in the inertial frame. {\it Panel (ii):} Damping times for the corotating branch, $\sigma^{s}_c$ is the mode frequency in the comoving frame.}
\label{fig:dampTimeCombined}
\end{figure*}

What is needed for the potentially unstable branch is a quantity that also changes its sign when a mode becomes prone to the CFS-instability and which is a monotonic function of the rotation rate. The $f$-mode frequency in the inertial frame $\sigma_i$ exactly conforms to these requirements. From equations \eqref{eq:equationsForFModesInCowling_I}, \eqref{eq:equationsForFModesInCowling_II} we can also make an estimation on how the damping times depend on the mode frequency $\sigma$. Since $dE/dt\sim \sigma^6$ and for any oscillation $E\sim\sigma^2$, we have
\begin{equation}
\label{eq:dampTimeUnstableEstimate}
\frac{1}{\tau}\sim \frac{dE/dt}{E}\sim\sigma^4\,.
\end{equation}
In Panel $(i)$ of Figure \ref{fig:dampTimeCombined}, we show the corresponding results of our simulations. There we plot normalized values of $(1/\tau)^{1/4}$ against normalized mode frequencies $\sigma_i$ in the inertial frame. Here, the normalization constants are the damping time $\tau_0$ and mode frequency $\sigma_0$ in the nonrotating limit. The rotation rate increases from right to left, i.~e.~with decreasing frequency, where the points $\mathcal{P}_1 = (1,1)$ and $\mathcal{P}_2 = (0,0)$ correspond to the nonrotating case and to an infinite damping time at the onset of the CFS-instability respectively.

%%%%%%%%%%%%% Figure %%%%%%%%%%%%%%%%%%%%%%%%
\noindent
%%%%%%%%%%%%% Figure %%%%%%%%%%%%%%%%%%%%%%%%

Since for a linear fitting, the two fixed points $\mathcal{P}_1$ and $\mathcal{P}_2$ already determine the fitting coefficients independent of our actual simulations, we decided to fit the data points with the next highest reasonbable polynomial which would be of cubic order. This not only fits the simulation data better but using a third order polynomial also turns out to be a very good approximation for the damping times of the corotating branches as we will see later.

A generic cubic polynomial is of the form $y(x) = ax^3 + bx^2 +cx +d$. Imposing the constraints, that the fit has to pass through $\mathcal{P}_1$ and $\mathcal{P}_2$ leads to $d = 0$ and $c = 1 - a - b$ and least-square methods finally yield
\begin{eqnarray}
\label{eq:fitForDampTimeUnstable}
\frac{\tau_0}{\tau} & = & \sgn{(\sigma^u_i)}\,0.256\left(\frac{\sigma^u_i}{\sigma_0}\right)^4\times\nn\\
& & \left[ 1 + 0.048\left(\frac{\sigma^u_i}{\sigma_0}\right) + 0.359\left(\frac{\sigma^u_i}{\sigma_0}\right)^2\right]^4,
\end{eqnarray}
where $\sgn{(x)}$ is the sign function. Equations \eqref{eq:inertialAndComovingFrequencyTransformation} and \eqref{eq:sigmaUnstableBranch} can then be used to cast this relation into a form that depends on the angular velocity again.

For the corotating branch, this type of scaling will not work for the following reasons. First, the damping times of the stable branch decrease with the rotation rate so instead of fitting $1/\tau$ we will rather use $\tau$ itself. Second and more importantly, the frequencies of the corotating branch do not depend monotonically on the rotation rate, see Panel $(ii)$ of Figure \ref{fig:splittingCombined}. For angular velocities near the mass-shedding limit, the mode frequencies tend to decrease. However, as already discussed in Section \ref{ssec:frequency_co_and_counter}, the order of the two nonaxisymmetric branches is reversed in the comoving frame. There, the frequencies of the comoving modes indeed again decrease monotonically with rotation rate, see once more Panel $(ii)$ of Figure \ref{fig:splittingCombined}. Thus, for these modes we can use the mode frequencies in the comoving frame as indicator for the angular velocity and the results for this type of parametrization is depicted in Panel $(ii)$ of Figure \ref{fig:dampTimeCombined}. Again, the rotation rate increases from right to left with $\mathcal{P}_1 = (1,1)$ representing the nonrotating limit.

Here, the spread of the data points in the $(\sigma,\tau)$-plane is larger when compared to the unstable branch but still it can be fitted very well with a third order polynomial. Especially the boosted decrease in the damping times for high rotation rates which directly correlates with the decrease of the mode frequencies in the inertial frame is captured very good with a cubic fit and cannot be reproduced properly by a quadratic poynomial.

Starting again with a generic cubic fitting function and including the point $\mathcal{P}_1$ leads to  
\begin{eqnarray}
\label{eq:fitForDampTimeStable}
\frac{\tau}{\tau_0} & = & -0.656\times\nn\\
& &\left[1 - 7.33\left(\frac{\sigma^s_c}{\sigma_0}\right) + 14.07\left(\frac{\sigma^s_c}{\sigma_0}\right)^2 - 9.26\left(\frac{\sigma^s_c}{\sigma_0}\right)^3\right]\,,\nn\\
& &
\end{eqnarray}
where equations \eqref{eq:inertialAndComovingFrequencyTransformation} and \eqref{eq:sigmaStableBranch} can be used to replace the comoving mode frequency by the rotation rate of the star.

Similar to Section \ref{ssec:frequency_co_and_counter}, an additional, model-independent relation for the damping time $\tau_0$ of nonrotating configurations is needed. As it was shown in \cite{1998MNRAS.299.1059A}, the behaviour of $R^4/(M^3\tau)$ with respect to the compactness $M/R$ proves to be quite insensitive to details of the particular equation of state. 

%%%%%%%%%%%%% Figure %%%%%%%%%%%%%%%%%%%%%%%%
\begin{figure}[htp!]
\centering
\includegraphics[width=0.48\textwidth]{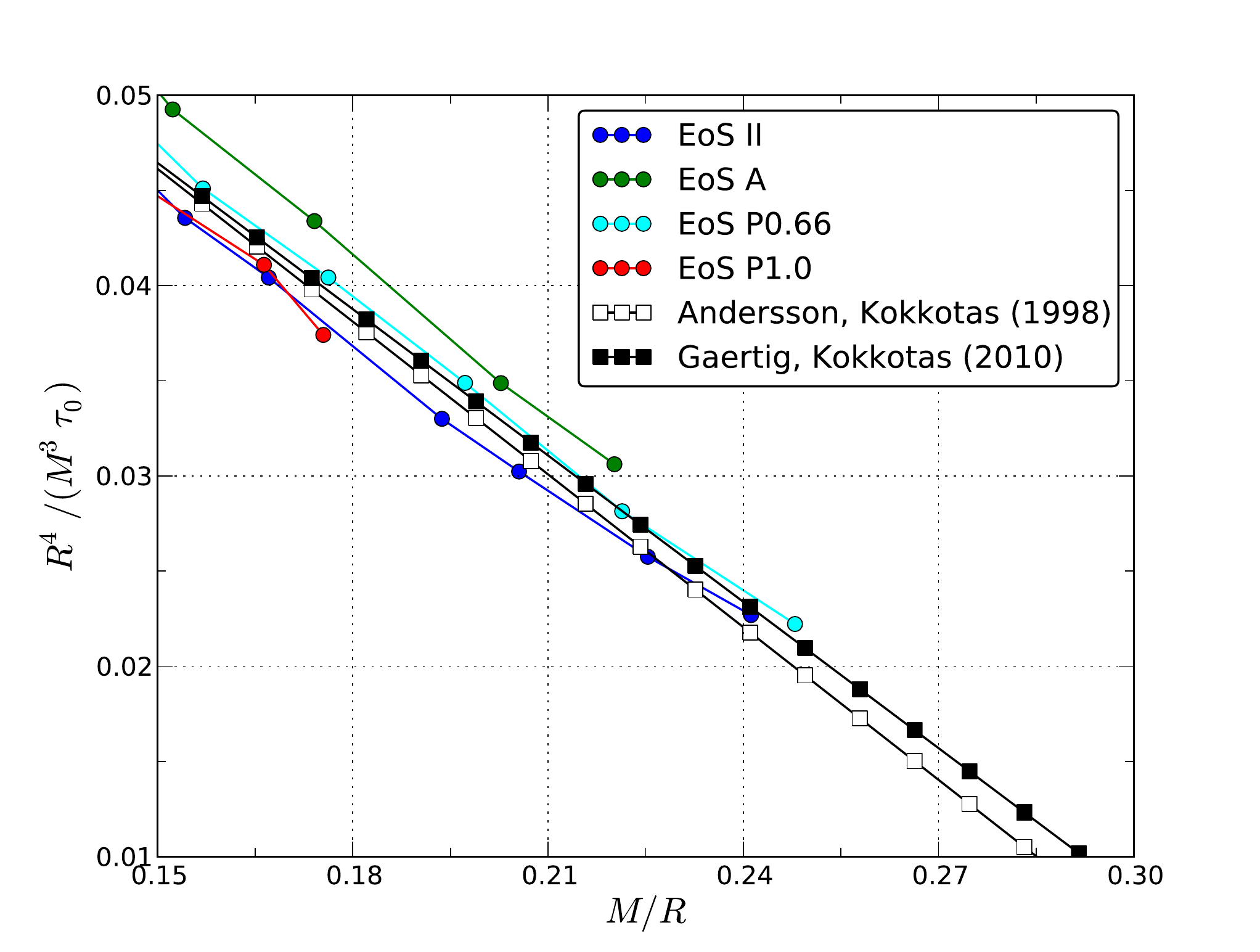}
\caption{Scaling of the $f$-mode damping time in the nonrotating limit. $M$, $R$ and $\tau_0$ are given in km.}
\label{fig:nonrotDampTime}
\end{figure}
\noindent
%%%%%%%%%%%%% Figure %%%%%%%%%%%%%%%%%%%%%%%%
In Figure \ref{fig:nonrotDampTime} we again compare our results with the corresponding findings in \cite{1998MNRAS.299.1059A}. In this case, only the softest equations of state from our sample, i.~e.~ EoS II, A, P0.66 and P1.0 attained high values in $M/R$ which allow for enough data points and a proper fitting in the compactness range depicted in Figure \ref{fig:nonrotDampTime}. As before, the overall behaviour is in good agreement with previous results; the larger spreading of the various EoS in Figure \ref{fig:nonrotDampTime} is most likely due to numerical errors which are introduced by the coordinate system used for our simulations.

Based on the data for nonrotating configurations, a linear fit leads to
\begin{equation}
\label{eq:tauZeroNonrot}
\frac{1}{\tau_0\,\mathrm{(s)}} = \frac{\bar{M}^3}{\bar{R}^4}\left[ 22.49 - 14.03\left(\frac{\bar M}{\bar R}\right)\right]\,,
\end{equation}
where again $\bar M$ and $\bar R$ are given by \eqref{eq:M_and_R_Normalized}.

The number of unknown variables which determine frequencies and damping times of the $f$-mode in relations \eqref{eq:sigmaStableBranch} - \eqref{eq:sigmaZeroNonrot} and \eqref{eq:fitForDampTimeUnstable} - \eqref{eq:tauZeroNonrot} can be reduced further by the well-known fact that to a very good accuracy, typically between 5 - 7\%, the Kepler-limit can be estimated by
\begin{equation}
\label{eq:empiricalMassShedding}
\Omega_K \approx 0.67\sqrt{\frac{GM}{R^3}}\,,
\end{equation}
see e.~g.~\cite{Friedman:1989lr,Haensel:1989fk}. This is only 23\% larger than the Newtonian value for polytropic stars. For realistic equations of state on the other hand, a similar empirical relation can be derived. In this case it was found that
\begin{equation}
\label{eq:empiricalMassShedding2}
\Omega_K = \mathcal{C}(\chi_s)\sqrt{\frac{GM}{R^3}}
\end{equation}
with
\begin{equation}
\label{eq:defOfChi}
\chi_s = \frac{2GM}{Rc^2}\qquad\mbox{and}\qquad\mathcal{C}(\chi_s) = 0.468 + 0.378\chi_s
\end{equation}
can reproduce the original values with a relative error of only 1.5\% \cite{1996ApJ...456..300L,2003LRR.....6....3S}. In this sense, $\Omega_K$ is not an independent parameter but can be computed very accurately from the mass and the radius of the nonrotating neutron star.

\subsection{Asteroseismology Examples}
\label{ssec:thought_experiment}
%=====================================================================

The empirical relations found in this Section can be used in two ways. By prescribing $M$, $R$ and $\Omega$, one can easily compute frequencies und damping times of both the co- and counterrotating mode branches for any rotation rate up the mass-shedding limit.

On the other hand they allow to do asteroseismology, for example three independent measurements, two frequencies and one damping time, will lead to a robust estimate of mass, radius and angular frequency and will therefore help to restrict the range of possible equations of state to those in agreement with these measurements. Of course, this is a very idealized point of view because it will be very difficult to observe damping times and frequencies of stable oscillations; this applies both for the co- and counterrotating modes. In these cases, a possible way of estimating the feasibility of a detection is to set a certain threshold on the gravitational wave amplitude and relate it to the energy that has to go into the $f$-mode as it was already done in \cite{1998MNRAS.299.1059A} for the nonrotating case. It was shown there, that the oscillations of a nascent neutron star can potentially be detected from within the local group of galaxies or even within the Virgo cluster if one assumes a more optimistic estimate for the radiated energy. 

Before damping/growth times are included in the examples, one might first ask what kind of information can be extracted by just detecting the $f$-mode frequencies of the co- and counterrotating branch. Since the relations \eqref{eq:sigmaStableBranch} - \eqref{eq:sigmaZeroNonrot} and \eqref{eq:empiricalMassShedding} depend on average density and rotation rate, one will not be able to determine mass and radius separately but merely the combination $M/R^3$ and $\Omega/\Omega_K$.

To give a simple example, we choose a certain background model of the less compact EoS II sequence with $f$-mode frequencies $\sigma_1/2\pi = 2.250$ kHz and $\sigma_2/2\pi = 1.804$ kHz in the inertial frame. Solving for the mean density and angular velocity yields $\bar{M}/\bar{R}^3 = 0.397$ and $\Omega/\Omega_K = 0.199$ while the correct values for this particular model are 0.467 and 0.173 respectively. As one can see, the relative error in average density and rotation rate is only around 15\% in both cases. Of course, in this example mass and radius cannot be determined independently; for this additional information about the damping times need to be taken into account as it is shown next.

We will address two separate questions here. The first one concerns the accuracy of the fittings when compared to the exact results and the second examines the possibility to use them for actual astroseismology, i.~e.~how accurate they constrain the neutron star parameters for a given set of measured frequencies and damping times.

As first example, we choose the less compact equilibrium sequence of EoS P0.66 and an arbitrary value for the rotation rate, e.~g.~$\Omega/2\pi = 0.676$ kHz which is roughly 38\% of the mass-shedding limit and corresponds to a ratio of polar to equatorial coordinate radius of $r_p/r_e = 0.95$. Inserting the values for mass and radius of this particular model from Table \ref{tab:backgroundModels} into relation  \eqref{eq:empiricalMassShedding} leads to $\Omega_K/2\pi\approx 1.742$ kHz which is only slightly smaller than the correct value of $\Omega_K /2\pi= 1.775$ kHz computed from the background code.    

The $f$-mode frequencies extracted from our time evolution are given by $\sigma_1/2\pi = 4.192$ kHz for the corotating and $\sigma_2/2\pi = 2.397$ kHz for the counterrotating branch; both frequencies are given in the inertial frame. On the other hand, evaluating the fitting formulas \eqref{eq:sigmaStableBranch}, \eqref{eq:sigmaUnstableBranch}, \eqref{eq:sigmaZeroNonrot}, \eqref{eq:empiricalMassShedding} with the correct values for $M$, $R$ and $\Omega$ yields $\tilde{\sigma}_1/2\pi = 4.205$ kHz and $\tilde{\sigma}_2/2\pi = 2.425$ kHz respectively which is an excellent match with the correct frequencies.

A similar comparison can be made for the damping times. Using the correct $f$-mode eigenfunctions for computing energy and energy-loss due to gravitational radiation, i.~e.~relations \eqref{eq:equationsForFModesInCowling_I} - \eqref{eq:equationsForFModesInCowling_III}, leads to $\tau_1 = 0.035$ secs and $\tau_2 = 0.559$ secs. Then again, when using the correct values for $M$, $R$ and $\Omega$ in relations \eqref{eq:sigmaStableBranch} - \eqref{eq:sigmaZeroNonrot} and \eqref{eq:fitForDampTimeUnstable} - \eqref{eq:empiricalMassShedding} one arrives at $\tilde{\tau}_1 = 0.050$ secs and $\tilde{\tau}_2 = 0.702$ secs respectively. This agrees quite well with the correct results; see Table \ref{tab:exampleEOS_C} for a summary.

%%%%%%%%%%%%% Table %%%%%%%%%%%%%%%%%%%%%%%%%%%%
\begin{table}[ht!]
\centering
\begin{tabular}{ccc}
\hline
parameter & exact value 	& value from the fit \\
\hline
$\Omega_K/2\pi$				&		1.775		&		1.742 \\
$\sigma_1/2\pi$				&		4.192		&		4.205 \\
$\sigma_2/2\pi$ 			&		2.397		&		2.425 \\
$\tau_1$					&		0.035		&		0.050 \\
$\tau_2$					&		0.559		&		0.702 \\
\hline
\end{tabular}
\caption{Comparison of the stellar parameters $\Omega_K$, $\sigma$ (both in kHz) and $\tau$ (in secs) for a less compact EoS P0.66 model with $\Omega = 4.247$ kHz between actual simulations and the empirical relations.}
\label{tab:exampleEOS_C}
\end{table}
%%%%%%%%%%%%% Table %%%%%%%%%%%%%%%%%%%%%%%%%%%%   
 
Overall, the accuracy by which the fitting relations can reproduce the exact results from actual simulations is reasonable. The difference is less than 2\% for the various frequencies and 25\% - 40\% for the damping times.
 
The second issue we would like to address is the inverse problem, i.~e.~how precise are the restrictions on mass, radius and angular velocity for a given triple of measurements. To continue with the previous example, we first look at the tuple of measurements $\mathcal{M}_1 := (\sigma_1, \sigma_2, \tau_1)$ that is the two $f$-mode frequencies of the co- and counterrotating branch and the damping time of the corotating mode. Reversing relations \eqref{eq:sigmaStableBranch}, \eqref{eq:sigmaUnstableBranch} and \eqref{eq:fitForDampTimeStable} leads to a system of nonlinear equations that is solved by an iterative root-finding algorithm as described in \cite{Powell:1970fk}. For a reasonable choice of starting parameters, this algorithm converges with adequate precision to an estimate of $M$, $R$ and $\Omega$. More specifically, from $\mathcal{S}_1$ we get $M = 1.44\,M_\odot$, $R = 8.98$ km and $\Omega/\Omega_K = 0.386$ and using relation \eqref{eq:empiricalMassShedding} with these values of mass and radius leads to $\Omega/2\pi = 0.669$ kHz.

One should note, that for a given tuple of measurements $\mathcal{M}$, the root-finding algorithm can in principle lead to other solutions as well, depending on the initial guess for the beginning of the iterations. For example, if one uses $(\bar{M}, \bar{R}, \Omega/\Omega_K) = (2.0, 1.0, 0.0)$ as starting values, the root finder converges to $M = 11.54\,M_\odot$, $R = 17.98$ km and $\Omega/\Omega_K = 0.386$. However, the estimates for $M$ and $R$ are well beyond the range of expected neutron star masses and radii so that they can safely be discarded although the rotation rate is matched perfectly. On the other hand, discarding these unphysical solutions, the nonlinear solver converges to essentially the same set of roots independently of the initial setup for the start of the iteration. In this example, the various solutions obtained with different initial guesses differ by less than 2\%.   

Alternatively, one can also check the corresponding results when providing the measurements $\mathcal{M}_2 := (\sigma_1, \sigma_2, \tau_2)$ as input data, this time with the damping time of the potentially CFS-unstable branch. Since now a different fitting function for $\tau$ is used, the corresponding findings from the nonlinear root solver will be slightly different in general. However, for this particular example we find practically the same values for $M$, $R$ and $\Omega/\Omega_K$ as in the first case; see Table \ref{tab:exampleEOS_C_2}.

%%%%%%%%%%%%% Table %%%%%%%%%%%%%%%%%%%%%%%%%%%%
\begin{table}[ht!]
\centering
\begin{tabular}{ccccc}
\hline
& $M$ & $R$ & $\Omega/2\pi$ & $\Omega_K/2\pi$ \\
\hline
exact 					& 1.10 	&	8.18 		& 0.676 	& 1.775\\
using $\mathcal{M}_1$ 	& 1.44	&	8.98		& 0.669		& 1.732\\
using $\mathcal{M}_2$ 	& 1.44	&	8.98		& 0.669		& 1.732\\
\hline
\end{tabular}
\caption{Comparison between exact and estimated stellar parameters from solving the inverse problem for measurements $\mathcal{M}_1$, $\mathcal{M}_2$. Here, $M$ is in units of $M_\odot$, $R$ in km and $\Omega$, $\Omega_K$ in kHz.}
\label{tab:exampleEOS_C_2}
\end{table}
%%%%%%%%%%%%% Table %%%%%%%%%%%%%%%%%%%%%%%%%%%% 

We repeated these two types of checks, i.~e.~comparison between exact results and fitting functions as well as solving the inverse problem, also for other EoS and larger angular velocities and a summary of these results for the more compact equilibrium model of EoS P1.2 is given in Table \ref{tab:exampleEOS_D}.

%%%%%%%%%%%%% Table %%%%%%%%%%%%%%%%%%%%%%%%%%%%
\begin{table}[ht!]
\centering
\begin{tabular}{ccc}
\hline
parameter & exact value 	& value from the fit \\
\hline
$\Omega_K/2\pi$				&		0.777		&		0.806 \\
$\sigma_1/2\pi$				&		2.259		&		2.283 \\
$\sigma_2/2\pi$ 			&		1.043		&		1.059\\
$\tau_1$					&		0.099		&		0.093 \\
$\tau_2$					&		4.716		&		6.650 \\
\hline
\end{tabular}
\begin{tabular}{ccccc}
&&&&\\
\hline
& $M$ & $R$ & $\Omega/2\pi$ & $\Omega_K/2\pi$ \\
\hline
exact 					& 1.58 	&	15.42 	& 0.464 	& 0.777\\
using $\mathcal{M}_1$ 	& 1.51	&	15.33	& 0.479		& 0.796\\
using $\mathcal{M}_2$ 	& 2.43	&	17.97	& 0.479		& 0.795\\
\hline
\end{tabular}
\caption{Comparison between exact and estimated stellar parameters for an EoS P1.2 model rotating at roughly 60\% of the mass-shedding limit. $M$ is in units of $M_\odot$, $R$ in km and $\Omega$, $\Omega_K$ in kHz while $\tau$ is given in secs.}
\label{tab:exampleEOS_D}
\end{table}
%%%%%%%%%%%%% Table %%%%%%%%%%%%%%%%%%%%%%%%%%%% 

This time, the angular velocity of the actual model is increased to about 60\% of the mass-shedding limit. Consequently, the damping time of the counterrotating mode $\tau_2$ is in the range of a few seconds already and it will continue to grow for more rapidly rotating models until the CFS-instability begins to operate. Still, the fitting functions provided in this study can reproduce the exact values for the frequencies within an error of 3\% while the difference in the damping times again is around 6\% - 40\%.

Concerning the solutions of the inverse problem, the two different measurements $\mathcal{M}_1$ and $\mathcal{M}_2$ lead to different values for mass and radius of the neutron star in this example. Here, using either $\mathcal{M}_1$ or $\mathcal{M}_2$ alone over- or understimates somewhat the correct values for $M$ and $R$ whereas the angular velocity and the Kepler-limit is matched almost perfectly with any of the datasets; the error there is around 3\%.

Altogether we were able to demonstrate the applicability of the fitting functions provided in this study for a wide range of different polytropic equations of state and rotation rates ranging from moderate to rapid. The accuracy of the empirical relations decreases for rapidly rotating models as expected but otherwise provides results within an error of a few percent for the frequencies. The dependency of the fits for the damping times on the mode-frequency is very strong; this is especially true for the potentially unstable modes and hence the estimates for the damping times are less accurate the closer one approaches the Kepler-limit. 

The solution of the inverse problem delivered reasonable values for stellar key parameters like mass, radius and rotation rate. Typically, using the measurements $\mathcal{M}_1$ or $\mathcal{M}_2$ alone either under- or overestimates the correct values for mass and radius while the rotation rate proved to be more robust to estimate by a single measurement alone. Ideally, the scheme presented here should be used iteratively in the following way: A first solution of the inverse problem already puts some constraints on the possible values for mass, radius and rotation rate in parameter space. Based on these constraints, certain EoS and rotation rates can be discarded and improved fits for frequencies and damping times within this region in parameter space will lead to further restrictions for mass, radius and angular velocity. This procedure can then be repeated until the desired accuracy is reached.

Another important issue is related to the issue of accurately determining mode parameters by an actual gravitational wave detector. However, a thorough discussion of measurement errors in frequencies and damping times is beyond the scope of this paper; we refer the interested reader to \cite{2001MNRAS.320..307K} and references therein.
  
%=====================================================================
\section{Summary}
\label{sec:summary}
%=====================================================================
In this work we demonstrated how one can do gravitational wave asteroseismology by using the frequencies and possibly the damping/growth times of the emitted waves from oscillating and rapidly rotating relativistic stars.  This is possible by the empirical relations that we have derived and which connect the frequencies and the damping/growth times of the oscillation modes with the stellar characteristics, i.e. with the mass, radius and rotation rate. We have actually shown that for polytropic equations of state of varying stiffness one can create very robust formulae connecting the observable frequency and damping times with the quantities like rotation frequency, average density and/or compactness. 

We have shown on a few examples how one can use the empirical formulae in order to derive the stellar parameters. In a realistic situation when an $f$-mode will be excited, it will be possible to detect the signal at least from galactic sources if the mode is CFS-stable and at least from sources in the Virgo cluster if it is unstable \cite{Lai:1995fk,Ou:2004qy,Shibata:2004uq,2010KWK}. This will be possible with the sensitivity of the advanced Virgo and LIGO detectors \cite{Acernese:2006fk,Abbott:2009qy} and probably even more feasible with the next generation gravitational wave telescopes such as ET (Einstein Telescope) \cite{2010CQGra..27h4007P,Andersson:2011lr}.

The ``weak'' point of the whole procedure relies in the approximate calculation of both frequencies and damping times. As we already mentioned we have neglected the spacetime perturbations and thus there is a systematic quantitative but not qualitative error in all data. Thus it is expected that the coefficients in relations \eqref{eq:sigmaStableBranch}, \eqref {eq:sigmaUnstableBranch} , \eqref{eq:fitForDampTimeUnstable} and \eqref{eq:fitForDampTimeStable} will be affected by a proper treatment of the spacetime degrees of freedom. However, it is believed that these changes will not alter the results significantly since the relations for frequencies and damping times are normalized with their corresponding values in the non-rotating limit. These values will absorb most of the differences when compared with the correct results in the presence of spacetime perturbations but of course, this is an issue that has to be addressed properly in future work. An additional outcome of this analysis will be the frequencies and damping times of the $w$-modes.

Finally, the empirical relations found in this study have been derived for polytropes of varying stiffness which are able to mimic the global properties of realistic equations of state; see for example EoS A and EoS II which are polytropic fits to tabulated EoS. Realistic hot equations of state are the best candidates for newly born neutron stars and have a higher chance of becoming unstable prior mutual friction completely suppresses any instability \cite{Andersson:2009lr} but currently our code is unable to perform time-evolutions of rapidly rotating neutron stars for generic tabulated data. Preliminary studies regarding this issue are promising but in a very early stage.
 
%=====================================================================
\section{Acknowledgements}
\label{sec:acknowledgments}
%=====================================================================
We are grateful to N.~Andersson, K.~Glampedakis, V.~Ferrari, L.~Gualtieri, E.~Berti and C.~Kr\"uger for critical reading of the manuscript, providing valuable suggestions for improvements.

This work was supported by the Deutsche Forschungsgemeinschaft (DFG) via SFB/TR7, E.~G.~ was funded by EGO via the VESF program.

\newpage

\appendix

%=====================================================================
\section{Numerical Procedure}
\label{sec:numerical_procedure}
%=====================================================================
This work is a continuation of our previous efforts \cite{2008PhRvD..78f4063G,2009PhRvD..80f4026G} and relies on the foundations laid therein. We will therefore briefly summarize the crucial parts of the previous studies that are needed for the computation of the damping times here.

We are numerically solving the relativistic hydrodynamics equations, linearized around background equilibrium configurations of uniformly rotating neutron stars. For this purpose, the time-evolution of the fluid perturbations is performed in a cylindrical coordinate frame $(\varrho, \zeta, \phi)$ which is comoving with the neutron star, surface-fitted for all rotation rates and where the metric takes the form 
\begin{eqnarray}
\label{eq:lineElement}
ds^{2}  & = & e^{-2U}\left[e^{2k}\left(d\varrho^{2}+d\zeta^{2}\right)+W^{2}d\varphi^{2}\right]\nn\\
 & & -e^{2U}(dt+a d\varphi)^{2}\,.
\end{eqnarray}
Here, the metric potentials $U$, $k$, $W$ and $a$ depend on $\varrho$ and $\zeta$ only and are obtained by solving the generalized TOV-equations for axisymmetric equilibrium configurations and a perfect-fluid energy-momentum--tensor. For computational purposes, the physical domain of the simulations
\begin{equation}
\label{eq:physDomain}
\mathcal{D} = [(\varrho, \zeta)\,, \varphi =  const.]
\end{equation}
is mapped onto a rectangular grid
\begin{eqnarray}
\label{eq:coordTransf}
\mathcal{T} & = & [s, t)\in [0,1]\times [0,2]\,,\nn\\
 & & (\varrho = \varrho(s, t), \zeta = \zeta(s, t)) \in \mathcal{D}]\,,
\end{eqnarray}
where the barotropic fluid equations for the perturbed velocity and pressure are discretized and, together with appropriate boundary conditions, are numerically intergrated by using an Iterated Crank-Nicholson scheme with an additional amount of Kreiss-Oliger dissipation.

Actually, we do not directly evolve the fluid perturbations but certain combinations of hydrodynamical and metric variables. This considerably reduces the complexity of the differential equations and simplifies the boundary treatment of the computational domain. More specifically, for azimuthal mode number $m$ our time-evolution variables are given by
\begin{eqnarray}
\label{eq:perturbedAnsatz}
f_{1}(\varrho,\zeta,t)e^{im\varphi} & = & (\epsilon + p)W e^{U}\,\delta u_{\varrho}\nn\\
f_{2}(\varrho,\zeta,t)e^{im\varphi} & = & (\epsilon + p)W e^{U}\,\delta u_{\zeta}\\
f_{3}(\varrho,\zeta,t)e^{im\varphi} & = & (\epsilon + p)\,\delta u_{\varphi}\nn\\
H(\varrho,\zeta,t)e^{im\varphi} & = & c_{s}^{2}e^{U}\,\delta\epsilon\nn \, ,
\end{eqnarray}
where $f_1$, $f_2$, $f_3$, $H$ are integrated in time and $\delta u_{\varrho}$, $\delta u_{\zeta}$, $\delta u_{\varphi}$, $\delta\epsilon$ are the perturbed fluid velocities and energy-density respectively. Furthermore, $p$ is the background pressure, $\epsilon$ the unperturbed energy-density and $c_s$ the speed of sound which can be computed analytically from the equilibrium configuration for polytropic equations of state.

After a successful time-integration, which is typically cancelled after 50 - 70 ms, one can use Fast Fourier Transforms to extract oscillation frequencies and eigenfunctions of any mode one wants to study. This data is then used for further post-processing; for example in order to compute damping times.

For this, one has to adapt equations \eqref{eq:equationsForFModesInCowling_I} - \eqref{eq:equationsForFModesInCowling_III} to the computational domain $\mathcal{T}$. The corresponding relations are
\begin{equation}
\label{eq:equationsInOurCoordinates_I}
E = \frac{1}{2}\int\sqrt{\gamma}\left[\rho\delta u^{a} \delta u^{\ast}_{a} + \frac{\delta p}{\rho}\delta\rho^{\ast} \right]|\text{det}\, J(s,t)|\,ds dt d\phi
\end{equation}
for the energy contained within a mode,  
\begin{equation}
\label{eq:equationsInOurCoordinates_II}
\frac{dE}{dt} = - \sigma_i(\sigma_i + m\Omega)N_2 \sigma_i^{4}|\delta D_{22}|^2
\end{equation}
for the energy-loss due to gravitational radiation and
\begin{equation}
\label{eq:massQuadrupole}
D_{22} = \int\sqrt{\gamma}\,\delta\rho\,r^2 Y^{\ast}_{22}|\text{det}\, J(s,t)|\,ds dt d\phi
\end{equation}
for the mass-quadrupole moment. Here
\begin{equation}
\label{eq:3_determinant}
\gamma = (W^2\exp(-2U) - a^2\exp(2U))\exp(-4U + 4k)
\end{equation}
is the determinant of the spatial 3-metric from \eqref{eq:lineElement}, $J(s,t)$ is the Jacobian matrix of the coordinate mapping \eqref{eq:coordTransf} and $r$ is the radial distance. Due to the azimuthal decomposition of the perturbation variables, the integration in $\phi$-direction is trivial and raising the covariant fluid perturbations with the inverse metric yields
\begin{eqnarray}
\label{eq:velocityTerms}
\delta u^{\varrho} \delta u_{\varrho} & = & \left(\frac{1}{(\epsilon + p) W\exp(k)}\right)^2 |f_1|^2\nn\\
\delta u^{\zeta} \delta u_{\zeta} & = & \left(\frac{1}{(\epsilon + p) W\exp(k)}\right)^2 |f_2|^2\\
\delta u^{\varphi} \delta u_{\varphi} & = & \left(\frac{\exp(U)}{(\epsilon + p) W}\right)^2 |f_3|^2\,.\nn
\end{eqnarray}
Note, that energy-density and pressure vanish at the stellar surface, the same applies for the metric potential $W$ along the rotation axis. These critical points might lead to numerical problems when evaluating the expressions in \eqref{eq:velocityTerms}. However, in practice it turns out that our time-integration ensures the correct behaviour of the perturbation variables in order for equations \eqref{eq:velocityTerms} to remain finite.

Furthermore, the adiabatic condition leads to
\begin{equation}
\label{eq:deltaP}
\delta p = \frac{H}{\exp(U)}
\end{equation}
for the pressure perturbation, whereas the corresponding change in density can be computed once an equation of state is specified. In this study, we are considering polytropic equations of state (EoS) which take the form
\begin{equation}
\label{eq:polytropicEOS}
p = K\rho^{1 + 1/N}\quad \text{where} \quad \epsilon = \rho + Np\,.
\end{equation}
Here $K$ is the polytropic constant, $N$ the polytropic exponent and $\Gamma = 1 + 1/N$ the polytropic index. In gravitational units ($G = c = M_\odot = 1)$, $K^{N/2}$ can be used as scaling factor and in this new system, one arrives at
\begin{equation}
\label{eq:deltaRho}
\delta\rho = \frac{1}{\Gamma}p^{1/\Gamma - 1}\delta p\,.
\end{equation}
Finally, the Jacobian matrix needs to be computed. This can either be done by first interpolating $\varrho(s,t)$, $\zeta(s,t)$ onto the computational domain $\mathcal{T}$, followed by a finite-difference routine or alternatively by interpolating the pseudo-spectral expressions for the transformation coefficients $a_{ij} := \{\partial x_i/\partial y_k\,;\, x_i \in (s,t)\,, y_k \in (\varrho, \zeta)$\} and using the inverse function theorem to obtain the Jacobian of the inverse function. Due to interpolation errors in the first approach, the subsequent finite-difference scheme leads to highly non-smooth results for the Jacobian $J$ and the error introduced there can be as high as $30\%$; we are therefore favouring the second approach.

Eventually, standard finite integration schemes are deployed to evaluate the integrals in the expressions for energy and energy-loss \eqref{eq:equationsInOurCoordinates_II} and \eqref{eq:massQuadrupole}.

%We will later discuss, how the shape of the $l = |m| = 2$ oscillation modes change with increasing angular velocity of the background star. Is is therefore essential to get an impression of how the extracted eigenfunctions are arranged on the computational domain. Figure \ref{fig:l2m2_dp} shows the power spectral density of the scalar perturbation variable $H$, see \eqref{eq:perturbedAnsatz} for the definition.

\begin{figure}[htp!] 
\centering
\includegraphics[width=0.48\textwidth]{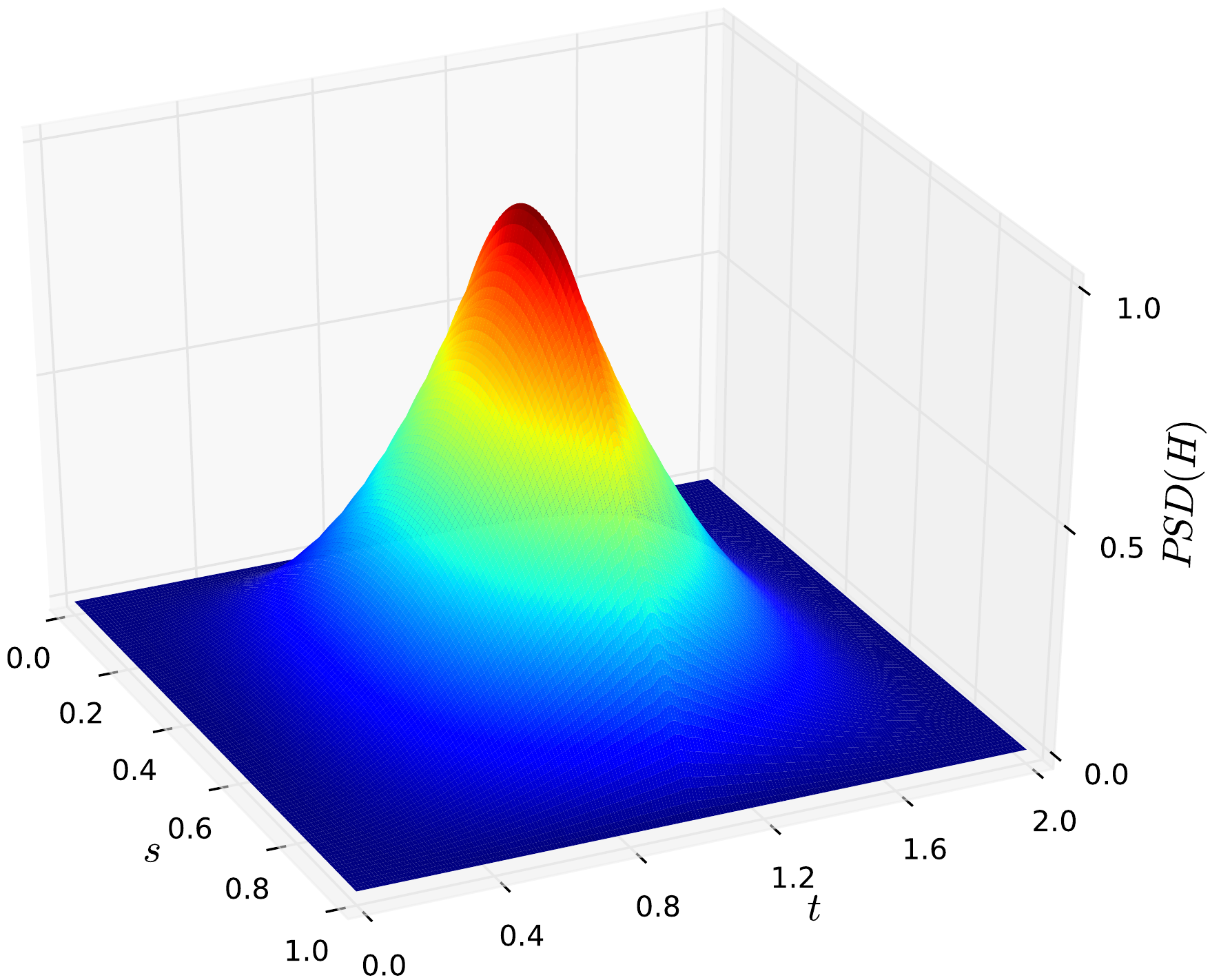}
\caption{Power spectral density of $H$ (normalized units) for the quadrupolar $f$-mode and a nonrotating background configuration.}
\label{fig:l2m2_dp}
\end{figure}

As an example and in order to get an impression of how the extracted eigenfunctions are arranged on the computational domain, Figure \ref{fig:l2m2_dp} shows the power spectral density of the scalar perturbation variable $H$, see \eqref{eq:perturbedAnsatz} for the definition.

There, the value of $s$ labels the radial coordinate which starts at the origin of the star for $s = 0$ and terminates at the stellar surface for $s = 1$. The use of surface-fitted coordinates ensure that the surface is always located at $s = 1$, even for rapidly rotating models. On the other hand, $t$ acts as an angular coordinate ranging from $t = 0$ at the rotation axis above the equatorial plane (i.~e.~$\zeta > 0$) to $t = 1$ at the equatorial plane and to $t = 2$ at the rotation axis below the equatorial plane ($\zeta < 0$).

%%%%%%%%%%%%% Figure %%%%%%%%%%%%%%%%%%%%%%%%
\noindent
%%%%%%%%%%%%% Figure %%%%%%%%%%%%%%%%%%%%%%%%
 
It is then clear, that the eigenfunction depicted in Figure \ref{fig:l2m2_dp} can indeed be identified with the fundamental $l = m = 2$ oscillation mode. It has no nodes in the radial direction and the angular pattern matches the scalar spherical harmonic $Y_{22}\sim\sin^2\theta$ which has a maximum for $\theta = \pi/2$, i.e. the equatorial plane, as well as global minima along the rotation axis. The eigenfunction also vanishes at the origin and along the surface of the star; this is due to boundary conditions and the special choice of the time-evolution variables, see \cite{2009PhRvD..80f4026G} for more details.

One should also keep in mind, that the power spectral density essentially measures the energy contained within a mode; this is at least true for the velocity perturbations $f_1$, $f_2$, $f_3$. Any auxiliary normalization coefficients that  are included from the Fast Fourier Transform-algorithms cancel out in the final equation \eqref{eq:dampingTime} for the damping time. This means, that we can directly use the data output of the eigenfunction extraction routine for computing the energy and energy-loss of the fundamental mode.

%=====================================================================
\section{Background Configurations}
\label{sec:background_configurations}
%=====================================================================
In this study, we treat neutron stars as perfect-fluid objects that obey a polytropic equation of state \eqref{eq:polytropicEOS}. Naturally, this is a rather crude approximation, neglecting a variety of micro-physical effects such as the true internal constitution of neutron stars (i.~e.~the distribution of baryons, leptons, optionally also hyperons and kaons) or the influence of a finite temperature as well as superfluidity and the existence of a solid crust (which will become important for temperatures around $10^{10}$ K), see \cite{Andersson:2007fk,Chamel:2008lr}. We also do not account for   the influence of magnetic fields which affect oscillation modes only for very high field strengths, see e.~g.~\cite{Lander:2010fk}.

In this sense, a simple polytropic equation of state parametrizes our ignorance about the true microphysical description of neutron star matter which is still unknown presently. However, one can neverthelsess use relativistic polytropes to cover the wide range of expected neutron star masses and radii \cite{Ozel:2006fk,Ozel:2008lr} and this is our proposed strategy here. We utilize a variety of polytropic EoS which are used in large parts in several other simulations of neutron star oscillations, either perturbatively or non-linear \cite{2008PhRvD..78f4063G,2009PhRvD..80f4026G,2010PhRvD..81h4019K,2004PhRvD..69h4008S,PhysRevD.81.084055,Font:2001qy,2006MNRAS.368.1609D,2004MNRAS.352.1089S,2006PhRvD..74f4024K,2008PhRvD..77f4019K}, two of them (EoS A and EoS II) actually are polytropic fits to tabulated data. Excluding this last two equations of state, the naming convention EoS P\# has been chosen here, where \# is replaced by the polytropic index $N$.

For each equation of state, we choose two different models; one rather close to the maximum allowed mass and another less compact one. A brief summary of basic stellar parameters for the non-rotating configurations can be found in Table \ref{tab:backgroundModels}.

%%%%%%%%%%%%% Table %%%%%%%%%%%%%%%%%%%%%%%%%%%%
\begin{table}[ht!]
\centering
\begin{tabular}{ccccc}
\hline
EoS 	& $K$ 	& $\Gamma$ 	& $M$ $(M_\odot)$ 	& $R_e$ (km) \\
\hline
II 			& 1186 	& 2.34	 	& 1.91 				& 11.62 \\
			&		&			& 1.50				& 13.19	\\
A 			& 1528 	& 2.46		& 1.61				& 9.47	\\
			&		&			& 1.25				& 10.54	\\
P0.66		& 1000	& 5/2		& 1.29				& 7.60	\\
			&		&			& 1.10				& 8.18	\\
P1.0			& 100	& 2.0		& 1.51				& 13.32	\\
			&		&			& 1.40				& 14.07	\\
P1.2			& 35		& 11/6		& 1.58				& 15.42	\\
			&		&			& 1.32				& 18.38 \\
P1.4			& 20		& 12/7		& 1.91				& 20.69	\\
			&		&			& 1.83				& 23.17	\\
\hline
\end{tabular}
\caption{EoS parameters and basic stellar properties for the nonrotating models. $K$ is given in dimensionless units $(G = c = M_\odot = 1)$, $M$ is the gravitational mass and $R_e$ the equatorial circumferential radius.}
\label{tab:backgroundModels}
\end{table}
%%%%%%%%%%%%% Table %%%%%%%%%%%%%%%%%%%%%%%%%%%%

Starting from the nonrotating configuration and by keeping the central rest-mass density fixed, the ratio of polar to equatorial coordinate radius $r_p/r_e$ is successively decreased, typically by a factor of $0.05$, until the mass-shedding limit is reached. In addition, supplementary equilibrium models were constructed especially for resolving the regime of slow rotation since for example moving from $r_p/r_e = 1.0$ to $r_p/r_e = 0.95$ for EoS P0.66 already means to reach roughly $40\%$ of the Kepler-limit.

Since the centrifugal force now supports the pressure in sustaining gravity, rotating models attain much larger masses. In Table \ref{tab:backgroundModelsMaxrot} we summarize fundamental stellar parameters for the maximally rotating equilibrium models. 

%%%%%%%%%%%%% Table %%%%%%%%%%%%%%%%%%%%%%%%%%%%
\begin{table}[ht!]
\centering
\begin{tabular}{ccccc}
\hline
EoS 			& $r_p/r_e$ 	& $\Omega_K/2\pi$ 	& $M$ $(M_\odot)$  	& $R_e$ (km) \\
\hline
II 			& 0.564 		& 1.393	 			& 2.61 (+37\%) 		& 15.64 (+35\%) \\
			& 0.546		& 0.998				& 1.94 (+29\%)		& 18.77 (+42\%)\\
A 			& 0.558 		& 1.759				& 1.97 (+22\%)		& 12.71 (+34\%)\\
			& 0.537		& 1.282				& 1.64 (+31\%)		& 15.05 (+43\%)\\
P0.66		& 0.555		& 2.180				& 1.58 (+22\%)		& 10.24 (+35\%)\\
			& 0.540		& 1.775				& 1.42 (+29\%)		& 11.50 (+41\%)\\
P1.0			& 0.579		& 0.966				& 1.80 (+19\%)		& 18.57 (+39\%)\\
			& 0.570		& 0.853				& 1.70 (+21\%)		& 19.83 (+41\%)\\
P1.2			& 0.597		& 0.777				& 1.81 (+15\%)		& 21.39 (+39\%)\\
			& 0.593		& 0.543				& 1.57 (+19\%)		& 26.19 (+42\%)\\
P1.4			& 0.62		& 0.538				& 2.13 (+12\%)		& 28.55 (+38\%)\\
			& 0.61		& 0.445				& 2.07 (+13\%)		& 32.64 (+41\%)\\
\hline
\end{tabular}
\caption{Basic stellar properties of the maximally rotating configuration and percental increase compared to the nonrotating case. $r_p/r_e$ is the ratio of polar to equatorial coordinate radius and $\Omega_K/2\pi$ represents the Kepler-limit, given in kHz.}
\label{tab:backgroundModelsMaxrot}
\end{table}
%%%%%%%%%%%%% Table %%%%%%%%%%%%%%%%%%%%%%%%%%%%
As it can be inferred from the data in this Table, the less compact models of any sequence can be deformed to a greater extent than the more compact ones as one would expect. This effect is of course more pronounced for softer equations of state.

\begin{figure}[ht!] 
\centering
\includegraphics[width=0.48\textwidth]{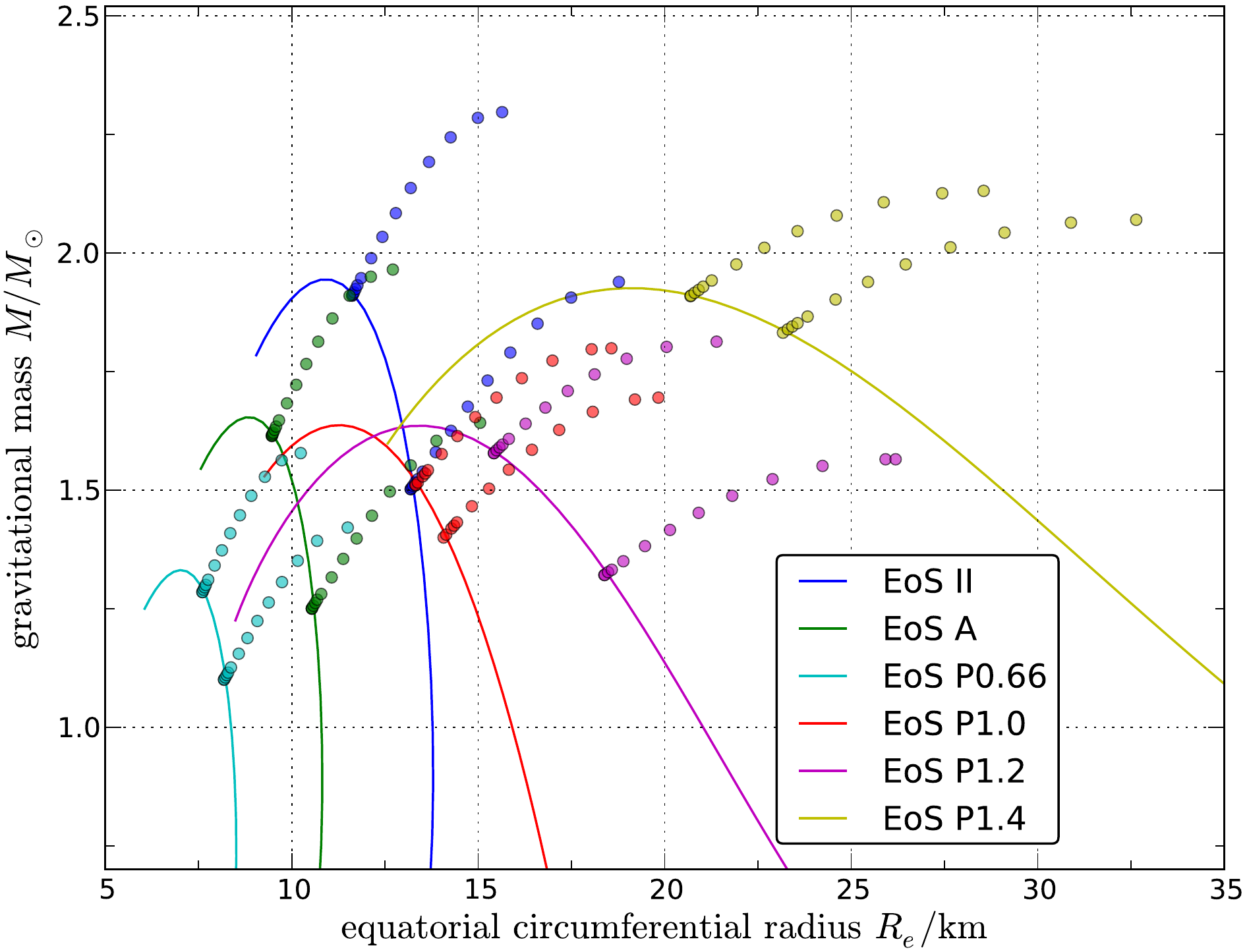}
\caption{Mass-Radius diagram for the six EoS and in our sample. The twelve different background sequences (two for every EoS) are marked as circles which branch off their corresponding non-rotating equilibrium curve.}
\label{fig:backgroundModels}
\end{figure}

In order of increasing stiffness, the equations of state investigated in this paper can be ordered $\text{P0.66} < \text{A} < \text{II} < \text{P1.0} < \text{P1.2} < \text{P1.4}$. This can also be seen from the mass-radius relations depicted in Figure \ref{fig:backgroundModels} where in addition the uniformly rotating background configurations are referenced as well.

%%%%%%%%%%%%% Figure %%%%%%%%%%%%%%%%%%%%%%%%
\noindent
%%%%%%%%%%%%% Figure %%%%%%%%%%%%%%%%%%%%%%%%

Typically, the increase in mass is not as pronounced as the increase in the radius when turning on rotation. While the percental increase in the gravitational mass can be as low as around $10\%$ in our sequences of constant rest-mass density, the corresponding increase in equatorial circumferential radius is always larger than at least $30\%$.

%=====================================================================
\section{Consistency Check}
\label{sec:consistency_check}
%=====================================================================
Here, we compare our results for the damping time of nonrotating polytropes with literature values in full general relativity.

In \cite{Balbinski:1985lr}, the authors compute damping times and fundamental mode frequencies for several relativistic polytropes with different polytropic indices. In Newtonian theory, damping times and mode frequencies for polytropes only depend on mass and radius of the neutron star and on the polytropic index $N$ of the particular equation of state but not on the polytropic constant $K$ or the central energy-density. This scale invariance is no longer valid in general relativity but it can be shown, that the relativistic calculations match with the Newtonian ones in the limit of small masses and large radii.

More specifically, if one defines a dimensionless parameter $c_\tau$ via
\begin{equation}
\label{eq:c_tau_definition}
\tau = c_\tau\frac{R}{c}\left(\frac{G M}{c^2 R}\right)^{-3}\,,
\end{equation}
then the relativistic results for a $\Gamma = 2$-polytrope approach the scale-invariant Newtonian value of $c_\tau = 8.46$ in the low-mass--limit.

For this purpose, a sequence of non-rotating background models was computed in \cite{Balbinski:1985lr} with $N =1$ and $K = 100$ km$^2$ and central rest-mass densities ranging from $\varrho_c = 3\times 10^{15} - 0.05\times 10^{15}$ g/cm$^3$. Then, damping times and oscillation frequencies of the fundamental quadrupolar mode were computed and compared with their Newtonian counterparts.

\begin{figure}[ht!] 
\centering
\includegraphics[width=0.48\textwidth]{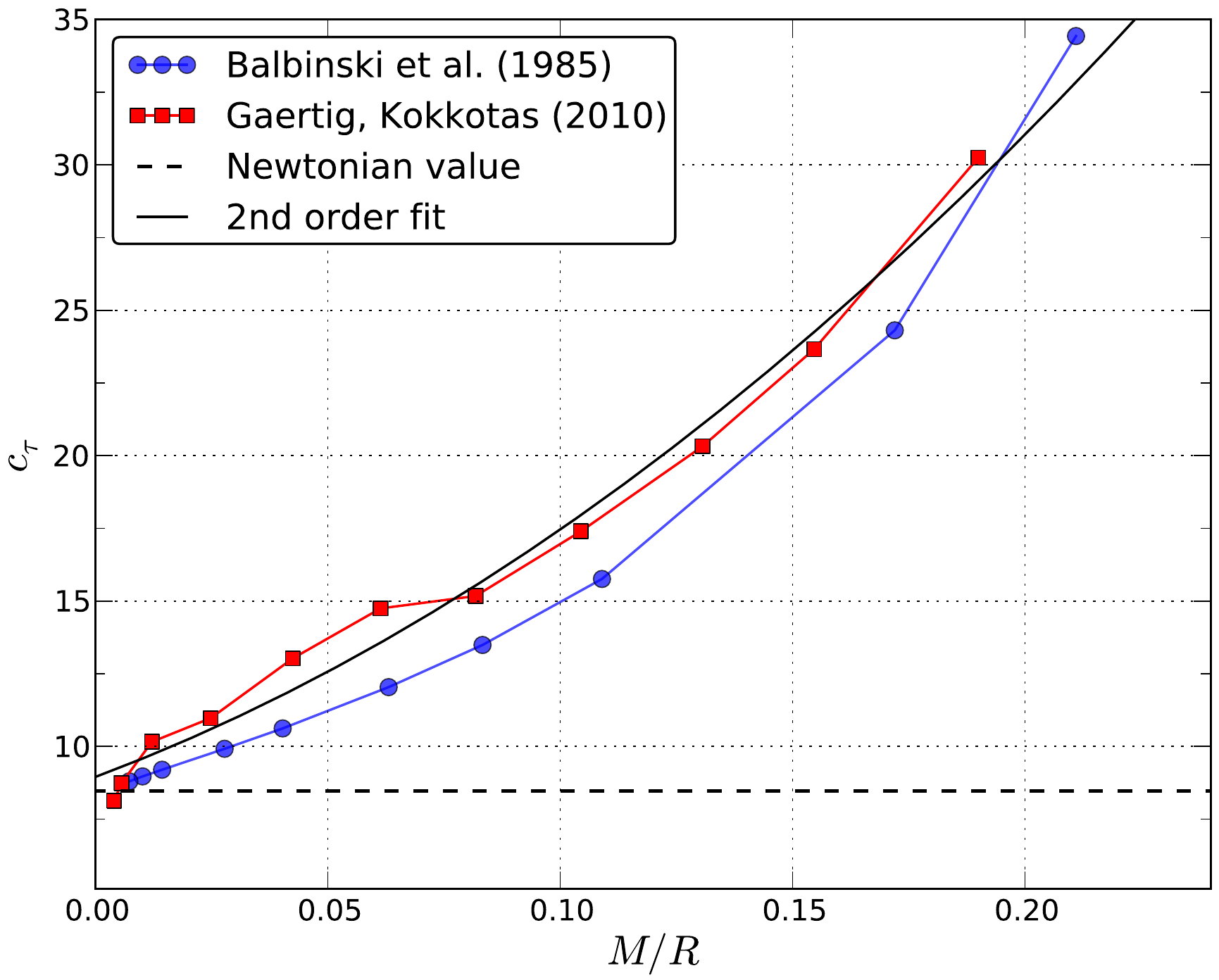}
\caption{Comparison between literature values and our results for $c_\tau$ (see \eqref{eq:c_tau_definition} for the definition). In the limit of small compactness, the Newtonian value of $c_\tau = 8.46$ is approached.}
\label{fig:balbinskiComparison1}
\end{figure}

We repeated the calculation with our code and an adjusted equilibrium sequence of EoS P1.0 which has the same polytropic index as the background models but a different value of $K$. Adopting $K = 100$ km$^2$ leads to more compact neutron stars with a maximum mass of $M_{max} = 1.1\,M_{\odot}$ and $R_{max} = 7.2$ km; compare with Figure \ref{fig:backgroundModels}. One should also keep in mind that the results in \cite{Balbinski:1985lr} are obtained by solving a complex eigenvalue problem. Here, we perform the calculation with our method of computing the eigenfunction from a time-evolution of the fluid perturbations and evaluating the integrals \eqref{eq:equationsInOurCoordinates_I}, \eqref{eq:equationsInOurCoordinates_II}. These are two completely different approaches and an agreement between these two methods would be a strong indication for the accuracy of our procedure. Figure \ref{fig:balbinskiComparison1} shows a comparison between the values of $c_\tau$ obtained by these two approaches.

%%%%%%%%%%%%% Figure %%%%%%%%%%%%%%%%%%%%%%%%
\noindent
%%%%%%%%%%%%% Figure %%%%%%%%%%%%%%%%%%%%%%%%
The overall agreement between the different methods is very good. Using a second-order fit for our results leads to a value of $c_\tau\approx 8.95$ in the Newtonian limit compared to $c_\tau = 8.46$ found in \cite{Balbinski:1985lr}. Especially for models with low $M/R$, the least compact object in our simulations has a mass of $M = 0.034\, M_{\odot}$ and a radius of $R = 12.5$ km, the $f$-mode frequencies are in the range of several hundred Hertz which impairs a proper eigenfunction extraction with our code.

The last remark concerns the limit of large masses in Figure \ref{fig:balbinskiComparison1}. Although we also use a polytropic equation of state, both are implemented slightly different here and in \cite{Balbinski:1985lr}. Here, a relation of the form \eqref{eq:polytropicEOS} is considered, which properly describes an ideal gas undergoing adiabatic processes, while the authors in \cite{Balbinski:1985lr} use $p = K\varrho^{1+1/N}$ with $\epsilon = \varrho$ in the relativistic case. This description of the fluid neglects the pressure contribution to the energy density and permits the speed of sound to become potentially larger than $c$ for all values of the polytropic index $N$, see also \cite{Tooper:1965lr}. Both prescriptions lead to the same stellar models in the limit of small compactness, because the energy density is dominated by the rest mass in this case and the contribution of the pressure is negligible. In more relativistic cases, the pressure contribution is noticeable and leads to a decrease in the mass of the most massive star which is dynamically stable to radial oscillations. This is the reason why our sequence of equilibrium configurations already terminates at $M_{max} \sim 1.1\,M_{\odot}$ while the sequence in \cite{Balbinski:1985lr} can reach up to $M_{max} \sim 1.3\,M_{\odot}$. However, this has only a modest effect on the computation of $c_\tau$.

\bibliographystyle{apsrev4-1}
\bibliography{references}

\end{document}